\documentclass[iop]{emulateapj}
\usepackage{morefloats}
\usepackage{epsfig}
\usepackage{epsf} 
	
\slugcomment{Submitted to The Astrophysical Journal}

\shorttitle{X-ZENS: X-ray point sources in ZENS groups at $z\sim0.05$}
\shortauthors{Silverman et al.}

\begin{document}


\title{The X-ray Zurich Environmental Study (X-ZENS). I. \\
$Chandra$ and XMM-$Newton$ observations of   AGNs in galaxies in nearby groups}


\author{J. D. Silverman \altaffilmark{1}, F. Miniati \altaffilmark{2}, A. Finoguenov \altaffilmark{3}, C.M. Carollo \altaffilmark{2}, A. Cibinel\altaffilmark{2,4}, S. J. Lilly \altaffilmark{2}, K. Schawinski \altaffilmark{2}}

\email{john.silverman@ipmu.jp}


\altaffiltext{1}{Kavli Institute for the Physics and Mathematics of the Universe, Todai Institutes for Advanced Study, the University of Tokyo, Kashiwa, Japan 277-8583 (Kavli IPMU, WPI)}
\altaffiltext{2}{Institute for Astronomy, ETH Z\"urich, CH-8093, Z\"urich, Switzerland.}
\altaffiltext{3}{Department of Physics, University of Helsinki, Gustaf H\"allstr\"omin katu 2a, FI-00014 Helsinki, Finland}
\altaffiltext{4}{CEA Saclay, DSM/Irfu/S\'ervice d'Astrophysique, Orme des Merisiers, F-91191 Gif-sur-Yvette Cedex, France}


\begin{abstract}

We describe X-ray observations with $Chandra$ and XMM-$Newton$  of 18 $M_{group}\sim1-6\times10^{13} M_{\odot}$, $z\sim0.05$ galaxy  groups from the {\it Zurich Environmental Study (ZENS)}. The X-ray data aim at establishing the  frequency and properties, unaffected by host galaxy dilution and obscuration, of AGNs in central and satellite galaxy members, also as a function of halo-centric distance.  X-ray point-source detections are reported for 22 of the 177 galaxies targeted by the X-ray observations, down to a sensitivity level of $f_{X}\sim 5\times10^{-15}$ erg cm$^{-2}$ s$^{-1}$ in the broad 0.5-8.0 keV band, corresponding to a limiting luminosity of $L_{0.5-8~{\rm keV}}\sim3\times10^{40}$ erg s$^{-1}$.  With the majority of the X-ray sources attributed to AGNs of low-to-moderate levels ($L/L_{Edd}\gtrsim10^{-4}$), we discuss the detection rate in the context of the occupation of AGNs to halos of this mass scale and redshift, and compare the structural and morphological properties between  AGN-active and non-active galaxies of different rank and location within the group halos.   We see a slight tendency for AGN hosts to have either relatively brighter/denser disks (or relatively fainter/diffuse bulges) than non-active galaxies of similar mass.  At galaxy mass scales $<10^{11} M_\odot$,  halo central galaxies  appear to be  a factor $\sim4$ more likely to host AGNs  than satellite galaxies of similar mass. This effect, coupled with the tendency for AGNs to be hosted by massive galaxies, explains the (weak) trend for AGNs to be preferentially found in the inner  parts of group halos, with no detectable trend with halo-centric distance  in the frequency of AGNs within the satellite population.  Finally, our data support other analyses in finding that the rate of decline with redshift of AGN activity in galaxy groups matches  that of the global AGN population, indicating that either AGN activity occurs preferentially in group  halos, or that the evolution rate is independent of halo mass.   These trends are of potential importance, and require X-ray coverage of a larger sample to be solidly confirmed.
\end{abstract}





\section{Introduction}

There has been much progress in recent years in recognizing the importance of both internal and external processes in shaping the properties of galaxies across cosmic time.  Both dynamical instabilities and feedback effects from either supernova or active galactic nuclei influence the stellar mass growth of individual galaxies.  With respect to external factors, there may be specific environments most conducive to the buildup of stellar mass, transformation of morphological type, and possibly black hole growth.  Galaxy group-scale dark matter halos with   masses of order $\sim10^{13}$ M$_{\odot}$ show a heightened population of bulge-dominated galaxies \citep{Wilman2009}, known to have enhanced levels of AGN activity \citep{Kauffmann2003,Pierce2007}.  This is possibly a result of gravitational tides and dynamical friction at their peak efficiency in accelerating galaxy evolution - and thus plausibly the growth of the  central super-massive black holes (SMBHs).  

Galaxy groups are indeed potentially sites where interactions and mergers occur on a cosmologically short timescale \citep[e.g.,][]{Barnes1990}.  Mergers are a credible candidate for triggering nuclear activity, given their ability to generate large mass inflow rates to the nuclear region thus fueling both starbursts \citep{Sanders1996} and AGNs \citep[e.g.][]{Hopkins2008}.  In fact, enhanced levels of AGN activity have been observed in close pairs of galaxies, both at low \citep{Ellison2011} and high \citep{Silverman2011} redshift, that are more common in galaxy overdensities similar to the group scale \citep{Lin2010,Kampczyk2013}.  Such a mechanism within these environments may lead to the quiescent black hole - bulge relation \citep{Gebhardt2000,Ferrarese2000} that is seen locally and may extend up to $z\sim1$ \citep{Jahnke2009,Schramm2013}, possibly indicative of a co-evolution scenario \citep[e.g.,][]{Merloni2004}.  

On the other hand, galaxies within  group halos may begin to be subject to forces that can cause gas depletion thus limiting or even shutting off star formation and the fueling of a SMBH \citep[e.g.,][]{Shin2012}.  For example, cold gas can be depleted through ram-pressure stripping, thermal evaporation or tidal stripping.  In fact, the stellar populations of galaxies in groups do show signs of a suppression of star formation that may be linked to their gas content \citep[e.g.,][]{vandenbosch2008,Peng2012,Knobel2013,Cibinel2012}.   

Groups at low redshift provide an environmental laboratory with sufficient spatial resolution to ask what physical processes actually impact galaxies that can then further initiate the fueling a of central black hole.  Several studies of AGN activity in nearby galaxy groups have been undertaken, but the statistics with respect to AGNs   and the selection of a well-defined parent group sample remains  a challenge \citep[e.g.,][]{Shen2007,Arnold2009}.  Recent large systematic redshift surveys of the local universe (2dfGRS: \cite{Colless2001}; SDSS: \cite{Yang2007}), with their well-defined catalogues of galaxy groups, offer the opportunity to conduct large efforts to study $z=0$ galaxies and their SMBHs in these potential wells \citep[e.g.,][]{Weinmann2006,Sabater2013}.    Our own Zurich ENvironmental Study (ZENS) focuses on multi-band optical data for a complete sample of galaxy groups extracted from the 2dFGRS, that enables an accurate parameterization of sub-structure in galaxies, including bulge-to-disk ratios, bar strengths, location and sizes of the star forming regions, strength of tidal features, etc \citep{Carollo2013, Cibinel2013,Cibinel2012}.  Therefore, ZENS offers an optimal sample to help clarify the physical drivers behind the coordinated growth of SMBH and their hosts in galaxy groups 

We have initiated a follow-up study of the X-ray emission from  ZENS groups (X-ZENS).  In this paper, we present the first-epoch X-ZENS observations of 18 groups observed with either $Chandra$ or XMM-$Newton$, and primarily describe the data  analysis and the point-like X-ray source population in these groups.  X-ray emission is a unique probe of low-luminosity, $L_X\lesssim 10^{40-42}$ erg s$^{-1}$ AGNs less affected by host galaxy dilution than optical tracers.  In a companion paper (Miniati et al. in preparation), we will describe the detection and properties of the thermal diffuse intra-group medium (DIM) emission from the ZENS groups,  a second key driver for our X-ZENS program.

By providing  a well-constrained  sample of low-luminosity AGNs, the currently available    X-ZENS data  give a first indication of (1)  the occupation frequency of AGNs in halos of mass $M_{group}\sim1-6\times10^{13} M_{\odot}$, i.e.,   intermediate-mass potentials within large-scale structure, (2)  whether AGNs show any preference to reside in galaxies of any given rank, i.e., central or  satellite, or at specific radial locations within these groups, and (3) compare the morphological and structural properties of the hosts of our AGNs to the overall ZENS galaxy population.  We highlight that this program is providing a local benchmark for higher redshift studies in key survey fields (e.g., COSMOS \citep{Scoville2007}, GOODS \citep{Giavalisco2004}, CANDELS \citep{Grogin2011}) that extend environmental studies of AGNs up to $z\sim2$ \citep{georgakakis2008,silverman2009a,smolcic2011,Tanaka2012a,Allevato2012}. Throughout this work, we assume $H_0=70$ km s$^{-1}$ Mpc$^{-1}$, $\Omega_{\Lambda}=0.7$, $\Omega_{\rm{M}}=0.3$, and AB magnitudes.
  
\begin{figure}
\epsscale{2}
\plottwo{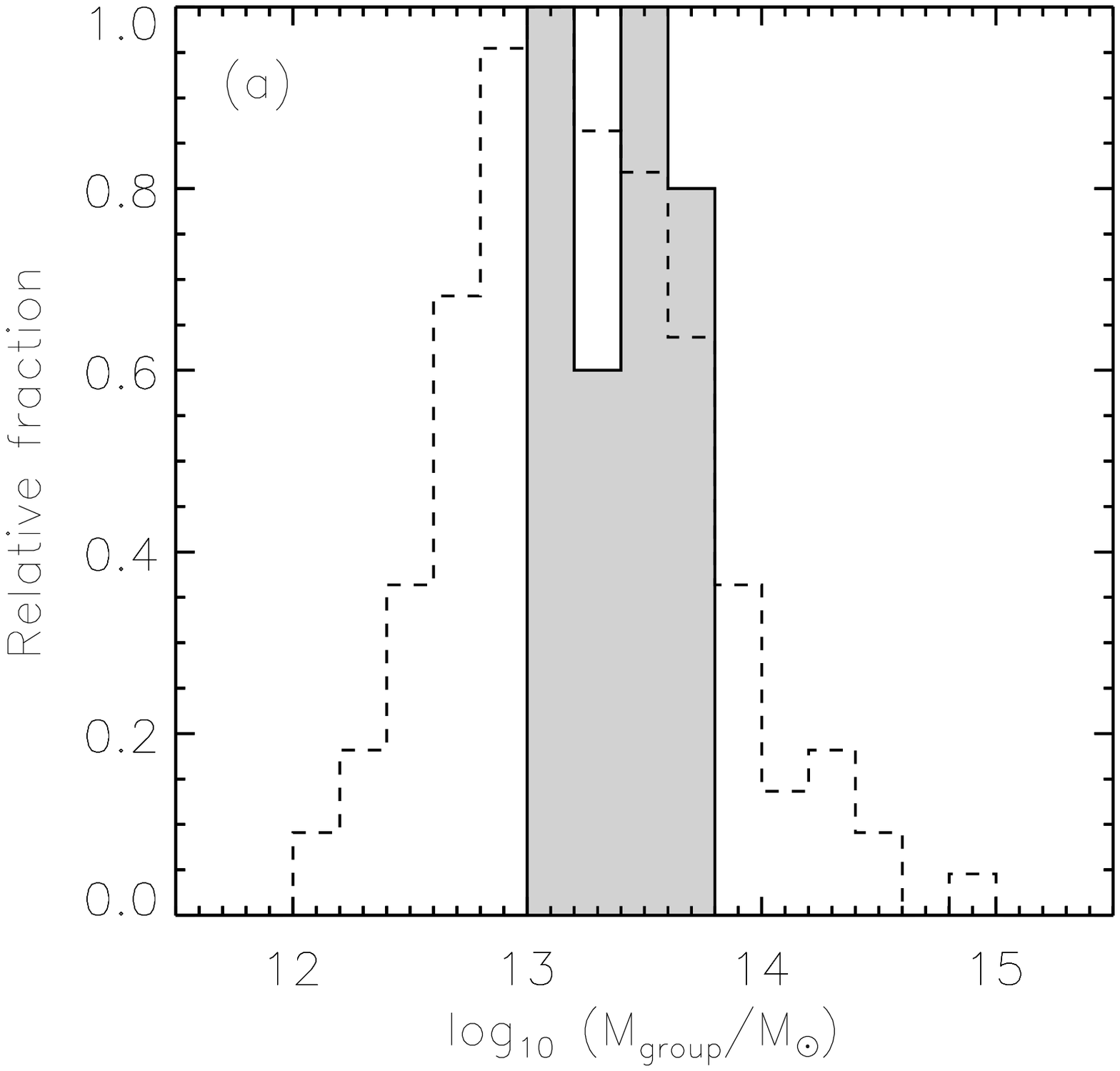}{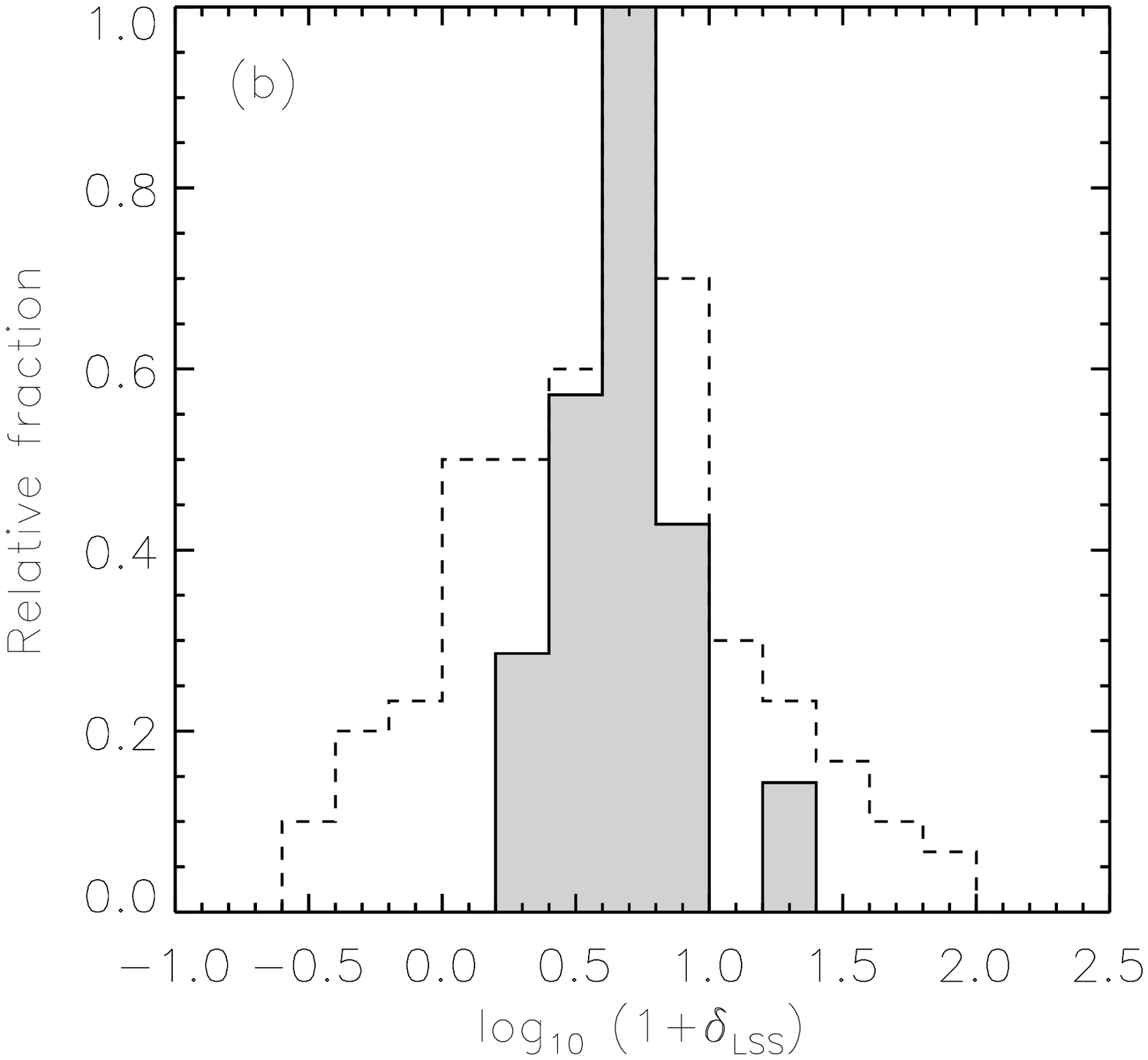}
\caption{Relative distribution of (a) halo mass of the group, and (b) the filamentary large-scale structure (LSS) density.  The full sample of 141 ZENS groups is shown by the dashed histogram while the filled histogram represents the 18 groups observed by either $Chandra$ or XMM-$Newton$.}
\label{group_prop}
\end{figure}

\begin{figure*}
\epsscale{0.8}
\plotone{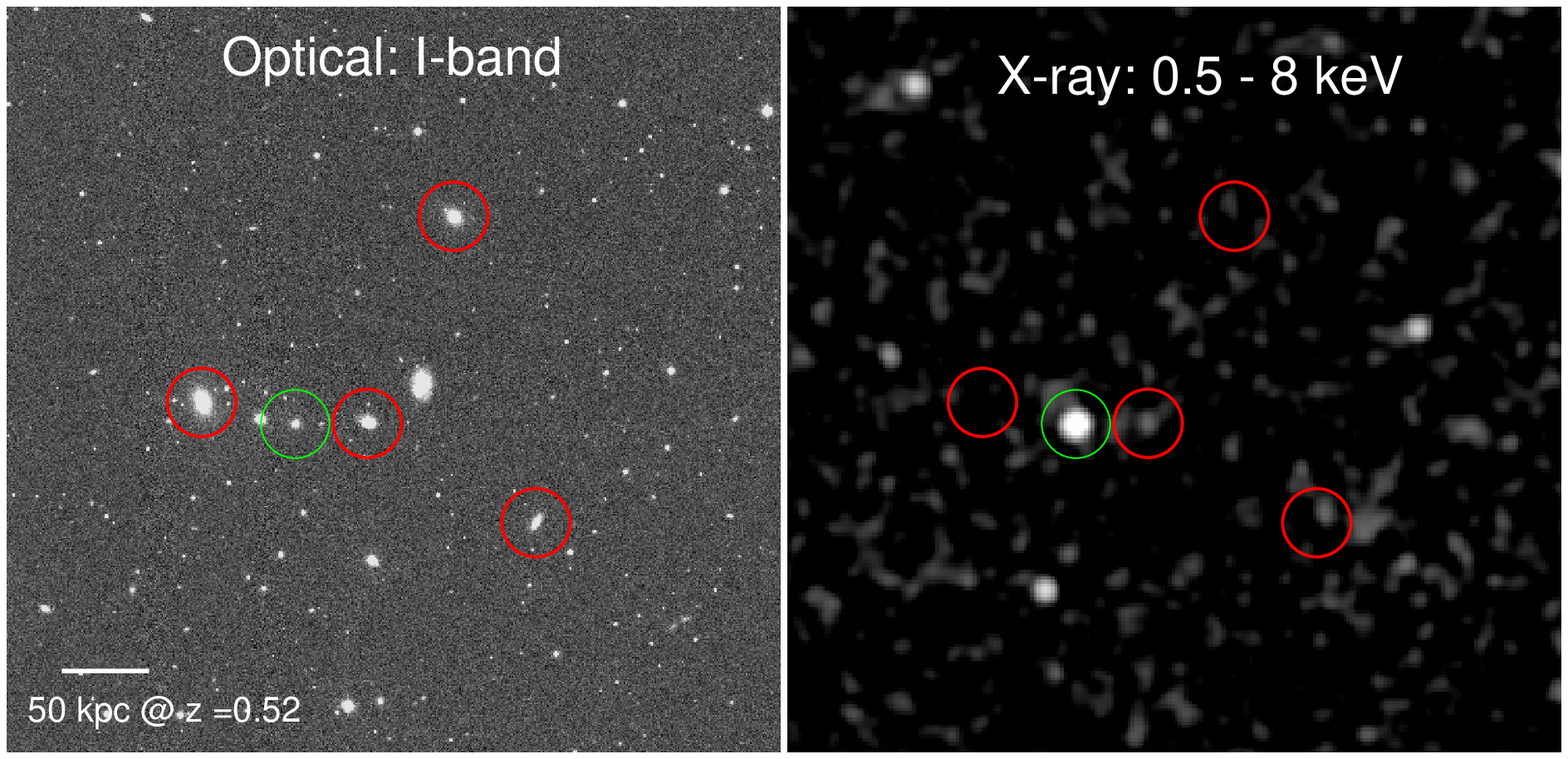}
\plotone{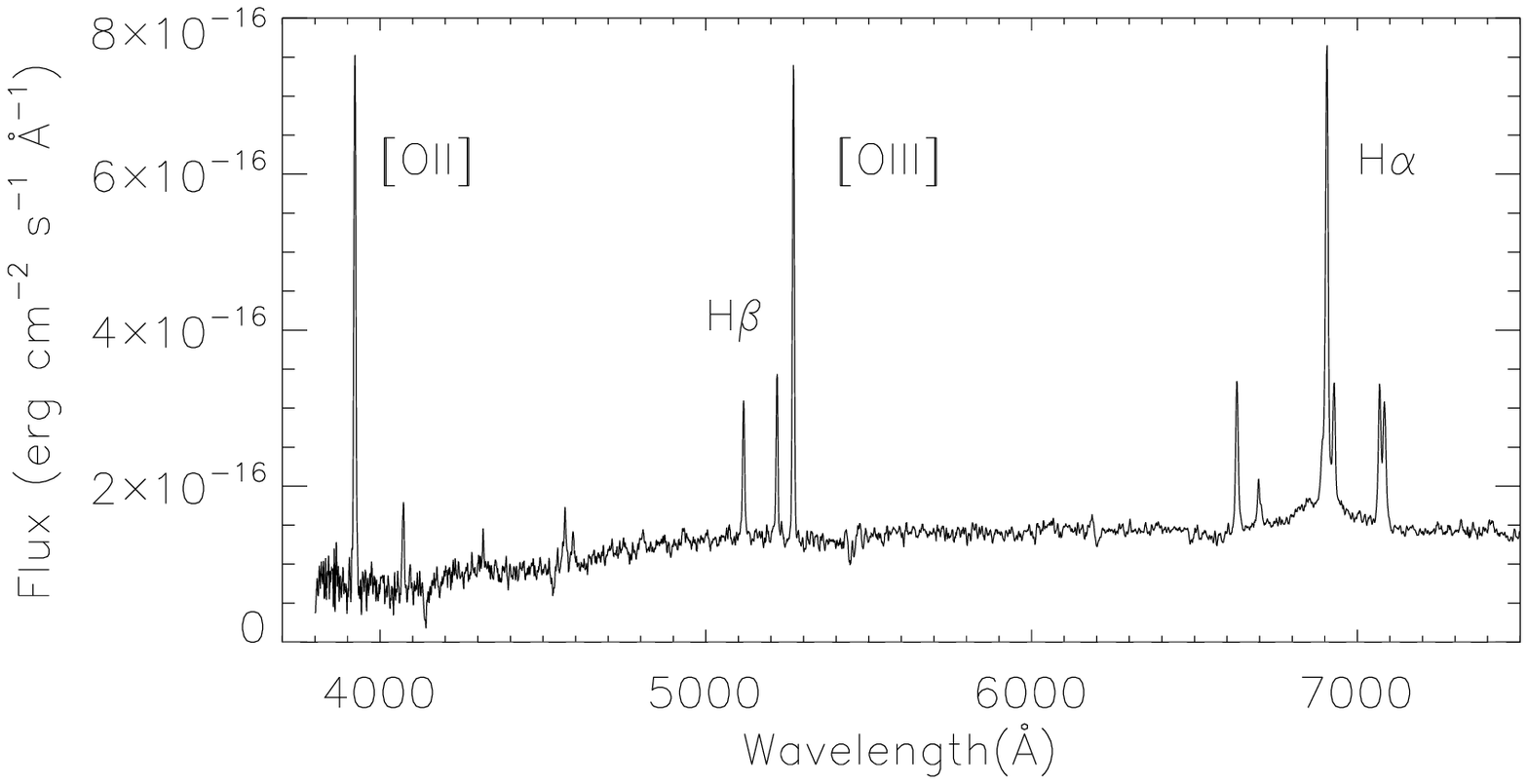}
\caption{SDSS J142808.89-023124.8.  Optical (I-band; {$\it left$ panel}) and $Chandra$ X-ray ($\it right$ panel) images of the central region of 2PIGG-n1381.  The X-ray image has been smoothed with a Gaussian (FWHM=2 pixels).  Galaxy group members are marked in both images by red circles ($r=20\arcsec$) with the central galaxy being the furthest to the left.  The optical and X-ray source associated with SDSS J142808.89-023124.8 is shown in green.  This object is clearly an AGN based on the broad emission component to the H$\alpha$ emission line sitting at its base.}
\label{fig:m1320}
\end{figure*}

\section{The Zurich Environmental Study (ZENS) of nearby galaxy groups}

ZENS\footnote{Online material related to ZENS can be found here http://www.astro.ethz.ch/research/Projects/ZENS} is a multi-wavelength study of galaxies in well-defined, optically-selected groups, primarily based on the 2dFGRS \citep{Colless2001,Colless2003}. This  large galaxy spectroscopic redshift survey has produced  nearly 225,000 galaxies at 14.5$<$b$_J$$<$ 19.5  with a median $<z> \sim 0.11$.  A friends-of-friends percolation algorithm was implemented to identify linked galaxies that likely share a common gravitational potential well; the resulting 2dFGRS Percolation-Inferred Galaxy Group (2PIGG) catalogue has 7000 groups \citep{Eke2004a,Eke2004b}.   

The structurally-resolved analyses of ZENS are based on deep, wide-field optical imaging, obtained at the ESO 2.2m telescope equipped with the WFI camera, in the $B$ and $I$ passbands. The ZENS  sample contains 141 groups extracted from the 2PIGG catalog \citep{Carollo2013,Cibinel2013,Cibinel2012}, for a total of 1630 galaxy members.   The  ZENS  sample was defined by selecting all groups within a narrow redshift range 0.05$<$z$<$0.0585 that have more than five galaxy members (N$_m$; number of constituent members in the 2dFGRS spectroscopic sample).  This ensures a very uniform group selection criteria and eliminates any distance related biases.  The $z$$\sim$0.05 redshift was chosen primarily because (1) the luminosities of the galaxies span $M_*$--$2$ to $M_*$+$3$, thereby covering all of the luminosity function of "massive" galaxies, (2) samples a range of group masses and large-scale environments, and (3) the ground-based resolution is well-suited to determine structural properties of galaxies such as bars, bulges, disks, tidal features and color gradients.   

For all galaxies in ZENS, we know the mass of the group halo in which they reside, their halo-centric (projected) distance, and their large-scale structure density; they are furthermore classified as centrals or satellites within their host group halos.  The group mass (M$_{group}$) is determined by the sum of the optical luminosity of each member (with a weighting scheme chosen to compensate for the survey incompleteness of the 2dFGRS) and assumed light-to-mass conversion factor.  The large-scale over-density ($\delta_{LSS}$) is measured by using the groups themselves as point-sources and measuring the distance to the 5th nearest neighbor; for this exercise, groups are weighted by their halo mass, restricted to be within a redshift interval $\delta z = \pm 0.01$ and required to be above a given optical luminosity.  Groups are further identified as relaxed or unrelaxed depending on whether a clear central galaxy is identified by considering both stellar mass, projected halo-centric distance, and relative velocity within the group.  We refer the reader to \citet{Carollo2013} for full details on the ZENS sample and derived physical properties.

\section{X-ray observations: data acquisition, analysis and source detection}

We extract from our ZENS survey a sample of 18 galaxy groups (Table~\ref{xray_sample}) for which we have acquired X-ray observations with either $Chandra$ or XMM-$Newton$ to primarily assess their AGN content by their X-ray emission and the presence of the thermal emission from the diffuse intragroup medium (DIM).  The point source sensitivity of $Chandra$ is ideal to detect AGNs to low luminosities  ($L_X\sim3\times10^{40}$ erg s$^{-1}$) with short exposure times (i.e., 10 ksec).  While the potential to detect extended emission from the DIM with $Chandra$ exists,  the higher collecting area in the soft-band of XMM-$Newton$ can provide more accurate measurements of spectral parameters (e.g., temperature) for the X-ray bright cases (Miniati et al. in preparation).  The groups for $Chandra$ followup are chosen to ensure that we sample the wide range in galactic composition by selecting those with (1) a number of spectroscopically-identified members $\ge$7, (2) a halo mass M$_{group}$ in the range $1.2-5.6\times10^{13}$ M$_{\odot}$,  and (3) a projected radii on the sky less then 9$\arcmin$ ($\hat{R}_{200}\lesssim 0.6$ Mpc) each, thus well matched to the $Chandra$/ACIS-I FOV (17$\arcmin$ $\times$17$\arcmin$).  Those observed by XMM-$Newton$ are similarly selected (see Miniati et al. in preparation for details).  Figure~\ref{group_prop} shows  the region of parameter space in both group mass (M$_{Group}$) and large-scale density ($\delta_{LSS}$) covered by our group sample with X-ray observations, in comparison with the full ZENS sample.    

\subsection{$Chandra$}

We  carried out $Chandra$/ACIS-I observations of 12 ZENS galaxy groups in Cycle 11 (PI J. Silverman; proposal \# 11700688; 120 ksec).  The observations were executed between September 2009 and October 2010.  Each target is observed for close to 10 ksec in order to detect AGNs at $z\sim0.05$ down to a limiting luminosity of $L_{(0.5-8~{\rm keV})} \sim 3\times 10^{40}$ erg s$^{-1} $ for detections with at least 4 counts in the broad energy band 0.5-8 keV.  The $16.9\arcmin \times 16.9\arcmin$ field-of-view (FOV) of CCDs $\#$0-3 of ACIS-I  is sufficient to cover the sky area of the target galaxy groups.  In total, there are 115 galaxies within the 12 ZENS groups that fall within the ACIS-I FOV.  The aim-points of $Chandra$ were chosen to maximize the number of galaxies that fall within the ACIS-I FOV.  This resulted in offsets as given in Table~\ref{chandra_log} from the group centers listed in Table~\ref{xray_sample}.  We also tried to avoid having galaxies falling within or near chip gaps; this was accomplished by adjusting the pointing location once the planned observation date was set thus the roll angle was known.  In other cases, this was needlessly achieved by splitting the observation into smaller time intervals and applying small offsets ($<1\arcmin$) to the aim point.  In Table~\ref{chandra_log}, we provide details on the individual exposures.              

We use CIAO tools (Version 4.3 and CALDB version 4.4.6) to perform the data analysis.  In cases where multiple exposures were taken, we use the task $merge\_all$ to combine the individual frames that generates a summed event file (level 2).  Source counts are measured in three energy bands (Broad (B): 0.5-8.0 keV, Soft (S): 0.5-2 keV, Hard (H): 2-8 keV) by placing circular apertures at the location of galaxies within our groups.  The extraction radius for each galaxy is set to include close to $100\%$ of the flux while the background contribution is estimated in an annulus centered on each individual galaxy with an area greater than the source region and free of any other nearby X-ray sources.  We consider a positive detection as any source with greater than or equal to 4 net counts in any of the three energy bands.  This is the same significance threshold employed by the Bootes survey \citep{Kenter2005} that has demonstrated that there is a low false-positive detection rate even at these low count levels.  We use the CIAO tool 'aprates' to estimate count rates and associated $1\sigma$ errors.  Exposure maps, required for conversion of count rates to physical fluxes (i.e., photons cm$^{-2}$ s$^{-1}$) , are generated in the three bands that correct for off-axis effects such as vignetting.  Due to the fact that $merge\_all$ does not properly deal with individual frames having a pointing offset, an exposure map was created for each individual  obs\_id ($mkinstmap$, $mkexpmap$), reprojected ($reproject\_image$) to a common astrometric frame and then coadded ($dm\_merge$).  These broad-band maps are based on a spectral weighting scheme that assumes a power-law spectral energy distribution with a photon index of 1.7.  Broad-band fluxes are then determined by multiplying the photon flux by the mean energy of a powerlaw distribution of photons with the spectral index as given above.  Errors on the total detected counts in the broad band allow an assessment of flux measurement errors while neglecting uncertainty of the assumed spectral model.  In Table~\ref{chandra_xray}, the measurements are provided for the individual $Chandra$ detections.  The magnitude of the error on the measured counts can be used to estimate uncertainties on flux and luminosity.  We note that there will be an additional error associated with the conversion of counts to energy units since we do not have full knowledge of the spectral shape of the individual detections due to the small number of counts.  

We further run wavdetect on each combined image in a given energy band.  This enables us to both validate the significance of the source detections with aperture estimates, as mentioned above, and to construct images devoid of X-ray point sources, required for the detection of diffuse emission (see Miniati et al. in preparation for details).  A significance threshold of $5\times10^{-6}$ was set to detect sources with low count statistics ($N\sim2-5$) while minimizing the inclusion of false detections.  The search was set to spatial scales of ${1, 2, 4, 8} \times 1.96\arcsec$ that are identical to those used by the $Chandra$ survey of the Bootes field \citep{Kenter2005}.

\begin{figure}
\includegraphics[angle=0,scale=0.45]{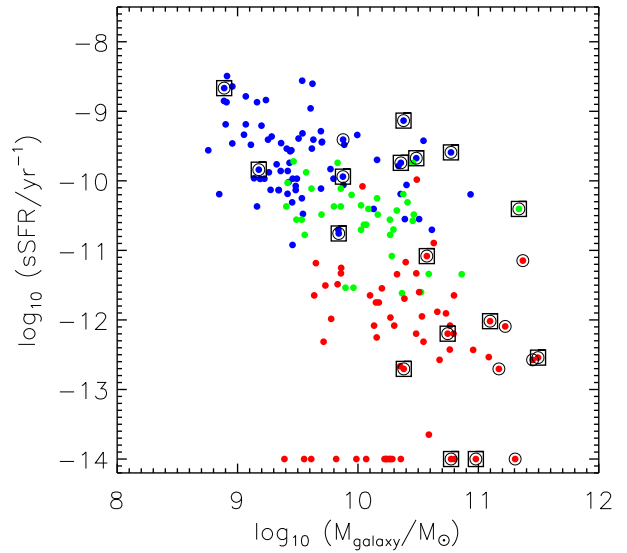}
\caption{Specific star formation rate versus stellar mass for galaxies in ZENS groups with X-ray coverage.  Color denotes three spectral types (quiescent - red, moderately star-forming - green, and strongly star forming - blue).  Galaxies firmly identified as quenched, but  with a large error in the sSFR obtained through template-fits to their SEDs, are all placed at a value of -14.  Galaxies with detected X-ray emission are identified with a large black circle; they cover the range of spectral types.  Open squares highlight ZENS galaxies with signs of AGN activity (either X-ray or optical).}
\label{gal_prop}
\end{figure}

\subsection{XMM-$Newton$}

We  also acquired X-ray imaging of 9 ZENS groups with XMM-$Newton$ (PI F. Miniati; proposal \#065530, \#067448, \#069374).  Of these, seven are of sufficient quality for our study of point-like emission as listed in Table~\ref{xray_sample} (flaring significantly degraded the quality of the data for the remaining two groups, thus rendering them as ineffective for our analysis).  The total exposure times for each observation, even after removing bad time intervals due to flaring, are sufficient to reach depths comparable to the $Chandra$ observations reported above.  This is due to the fact that the XMM observations were taken with the aim of providing sufficient count statistics  to accurately measure the physical properties of the DIM.  While four additional groups have some archival X-ray coverage with XMM-$Newton$, they are highly offset from the center position of our groups and miss most of the galaxy members.  One of them (2PIGG\_n1377) suffers from flaring and is not useful even though a single bright X-ray source is present that is associated with a galaxy (2PIGG\_n1377\_18) belonging to the group.  
Out of these seven groups, only one (2PIGG\_s1571) is in common with the $Chandra$ sample.  In fact, this group clearly has diffuse emission detected with both instruments as presented in Miniati et al. (in preparation).  The XMM-$Newton$ observation further aids in the detection of a X-ray point source associated with the central galaxy in 2PIGG\_s1571 that was uncertain from the $Chandra$ observation.  This highlights the complementarity of $Chandra$ and XMM.    

For point-source detection, we use the 0.5-2 keV and 2-7.5 keV bands while masking those energy intervals impacted by strong instrumental effects, as in \citet{Finoguenov2007}. A detailed modeling of the unresolved background, foreground and detector background has been undertaken, following the prescription of \citet{Bielby2010}.  We use $4\arcsec$ pixels and run the wavelet image reconstruction on 8$\arcsec$ and 16$\arcsec$ scales, with a 4$\sigma$ detection threshold.  We produce catalogs, based on the detections in two energy bands.  Physical flux units are obtained by converting the count rate (adjusted to account the flux losses within the 15" flux extraction circle due to XMM PSF) using PIMMS with a power law photon index of $\Gamma=1.7$, and no correction for intrinsic absorption.  We then cross-matched our X-ray catalog to the positions of the ZENS galaxies.  X-ray luminosities in the broad-band 0.5-8.0 keV are used throughout.  In Table~\ref{xmm_xray}, the measurements are provided for XMM-$Newton$ point-source detections.  As for the $Chandra$ sources, the errors on flux and luminosity can be propagated from the error on count rate.  In total, we have X-ray coverage of 177 unique galaxies that fall within either the $Chandra$ or XMM footprints.    

\section{X-ray emission from ZENS galaxies}

\label{pointsources}

\begin{figure}
\epsscale{2}
\plottwo{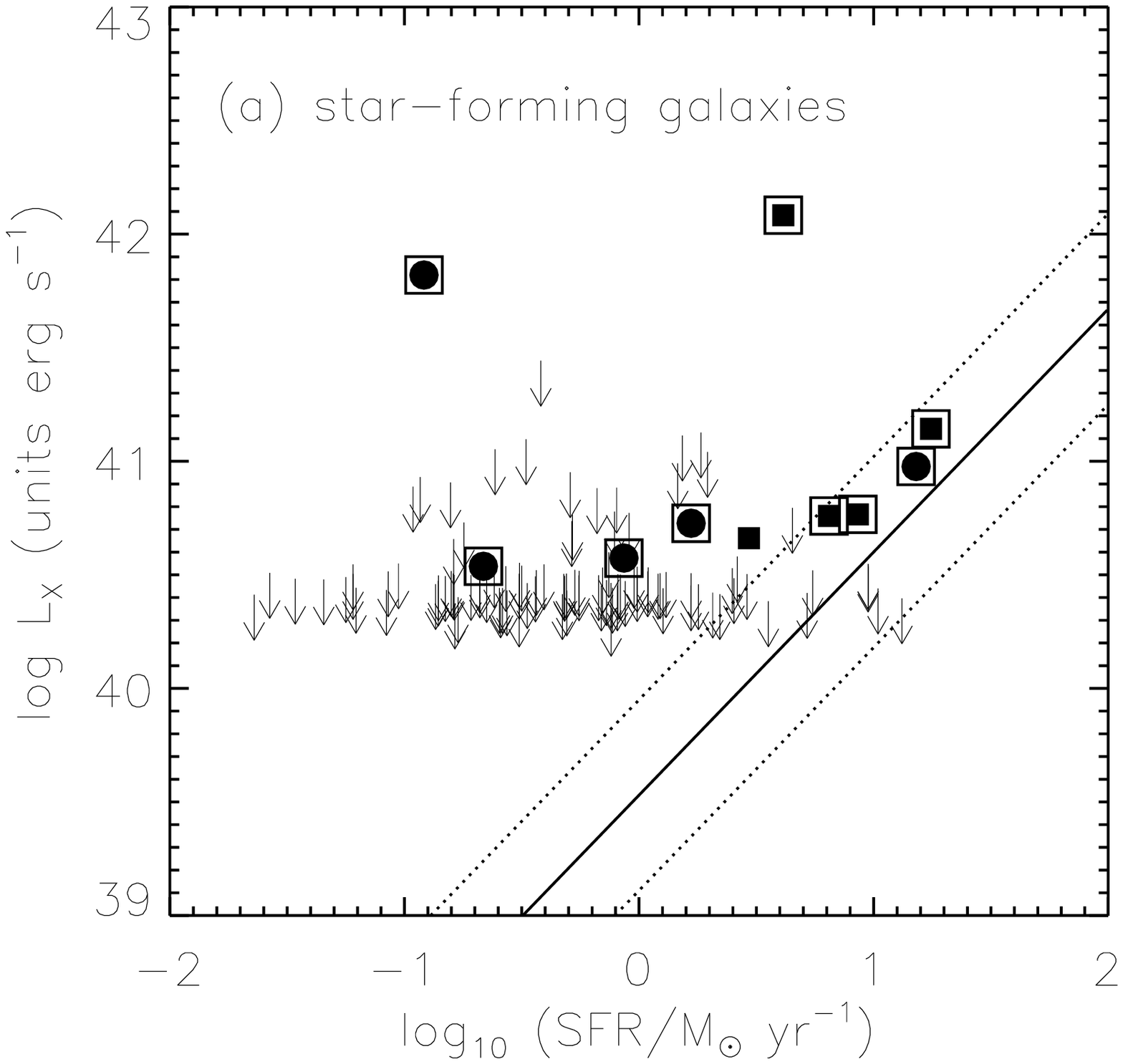}{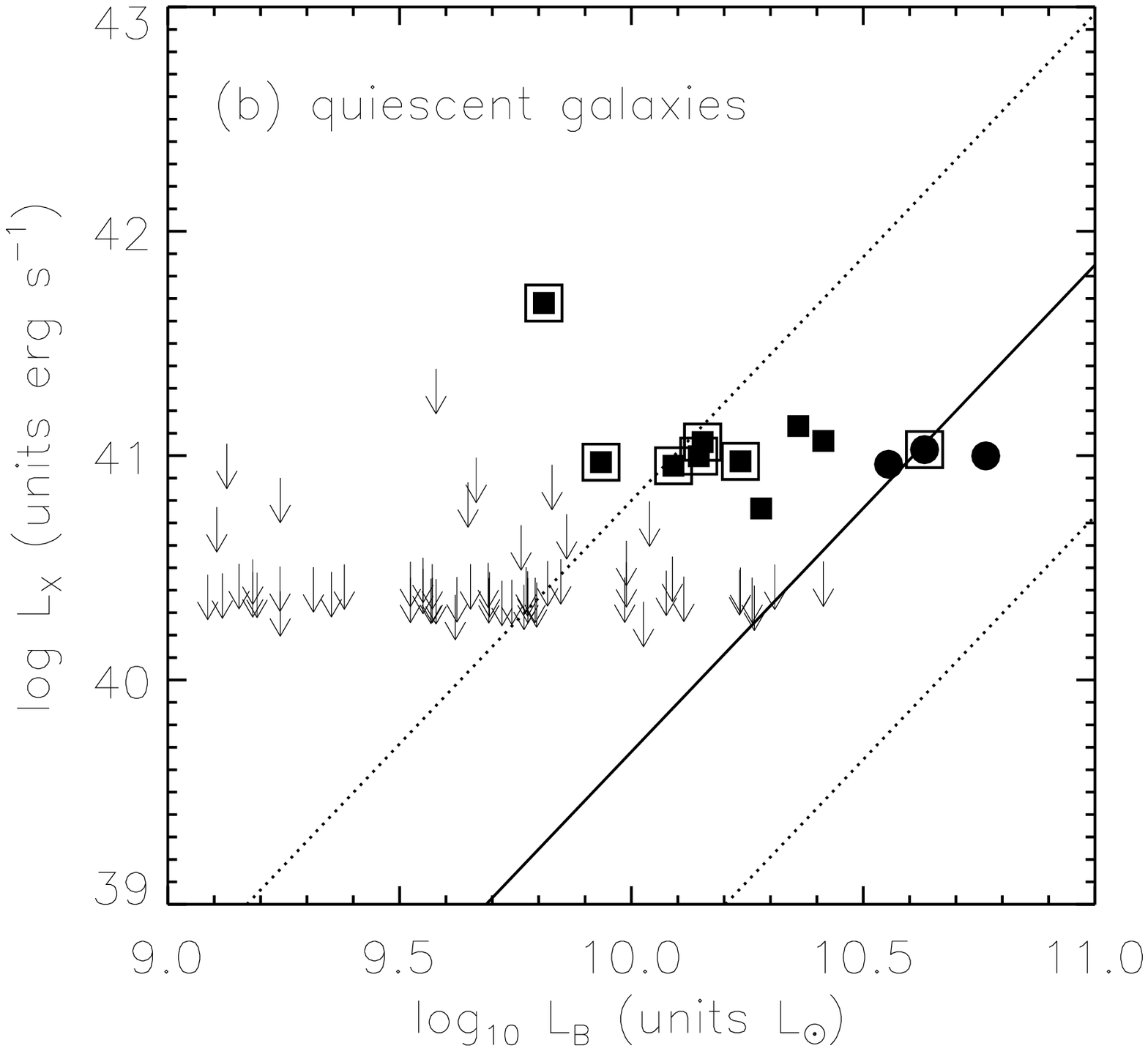}
\caption{X-ray luminosity for point-like sources split into star-forming ($a$) and quiescent ($b$) galaxies.  Black points mark individual detections ($Chandra$: filled circles; XMM-$Newton$: filled squares) while arrows indicate upper limits.  Slanted lines denote the region where normal galaxies lie either star-forming or quiescent with a 1$\sigma$ interval indicated by the dashed lines.  Open squares highlight ZENS galaxies with signs of AGN activity (either X-ray or optical).}
\label{xtype}
\end{figure}

Out of the 177 galaxies that are confirmed members of the 18 groups observed by $Chandra$ and XMM-$Newton$, we detect X-ray emission from 22 galaxies with at least 4 counts in one of the three $Chandra$ X-ray bands (broad: 0.5-8 keV, soft: 0.5-2.0 keV or hard: 2-8 keV) or above a S/N of 4 in either the soft (0.5-2 keV) or hard (2-7.5 keV) bands with XMM-$Newton$.  Pertaining to the $Chandra$ observations, the on-axis detection limit for our sample is $\gtrsim 5\times10^{-15}$ erg cm$^{-2}$ s$^{-1}$ that corresponds to a limiting X-ray luminosity of  $\sim3\times10^{40}$ erg s$^{-1}$ at the redshift $z\sim0.05$ of our group sample.  The flux limits of the XMM-$Newton$ observations are comparable as demonstrated by their typical luminosity limit of a $\sim5\times10^{40}$ erg s$^{-1}$.  All X-ray source detections have fluxes between 10$^{-15}$ and $10^{-13}$ erg cm$^{-2}$ s$^{-1}$ that correspond to a luminosity range above the limiting value given above and a maximum luminosity of 7$\times10^{41}$ erg s$^{-1}$.  The optical properties of these individual X-ray detections are provided in Tables~\ref{opt_table1} and~\ref{opt_table2}.  All optical properties are reported as given in the ZENS catalog \citep{Carollo2013, Cibinel2013, Cibinel2012}.

\begin{figure*}
\vskip 1cm
\hskip 0.3cm \includegraphics[width=4.7cm]{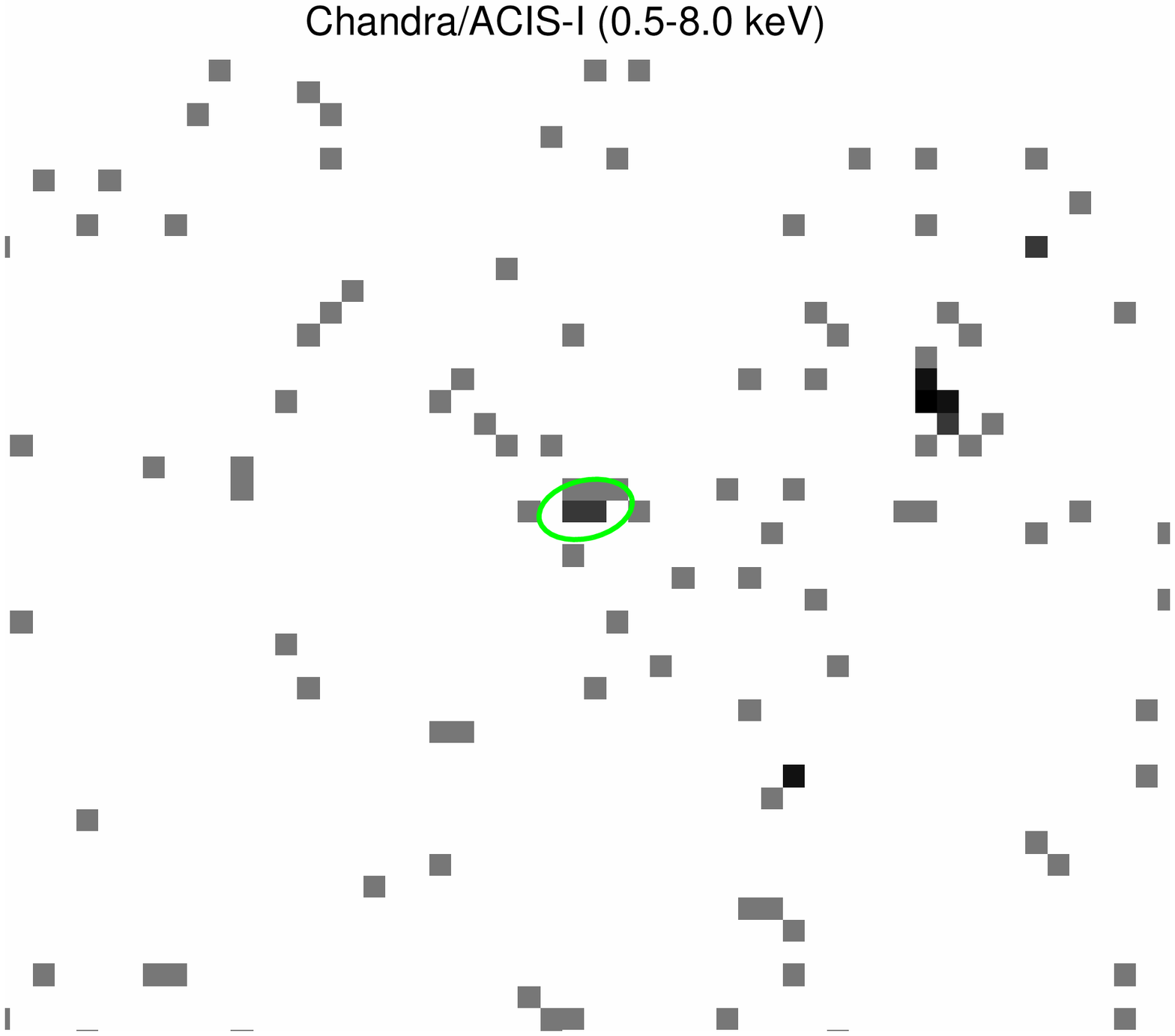}
\hskip 0.8cm \includegraphics[width=4.7cm]{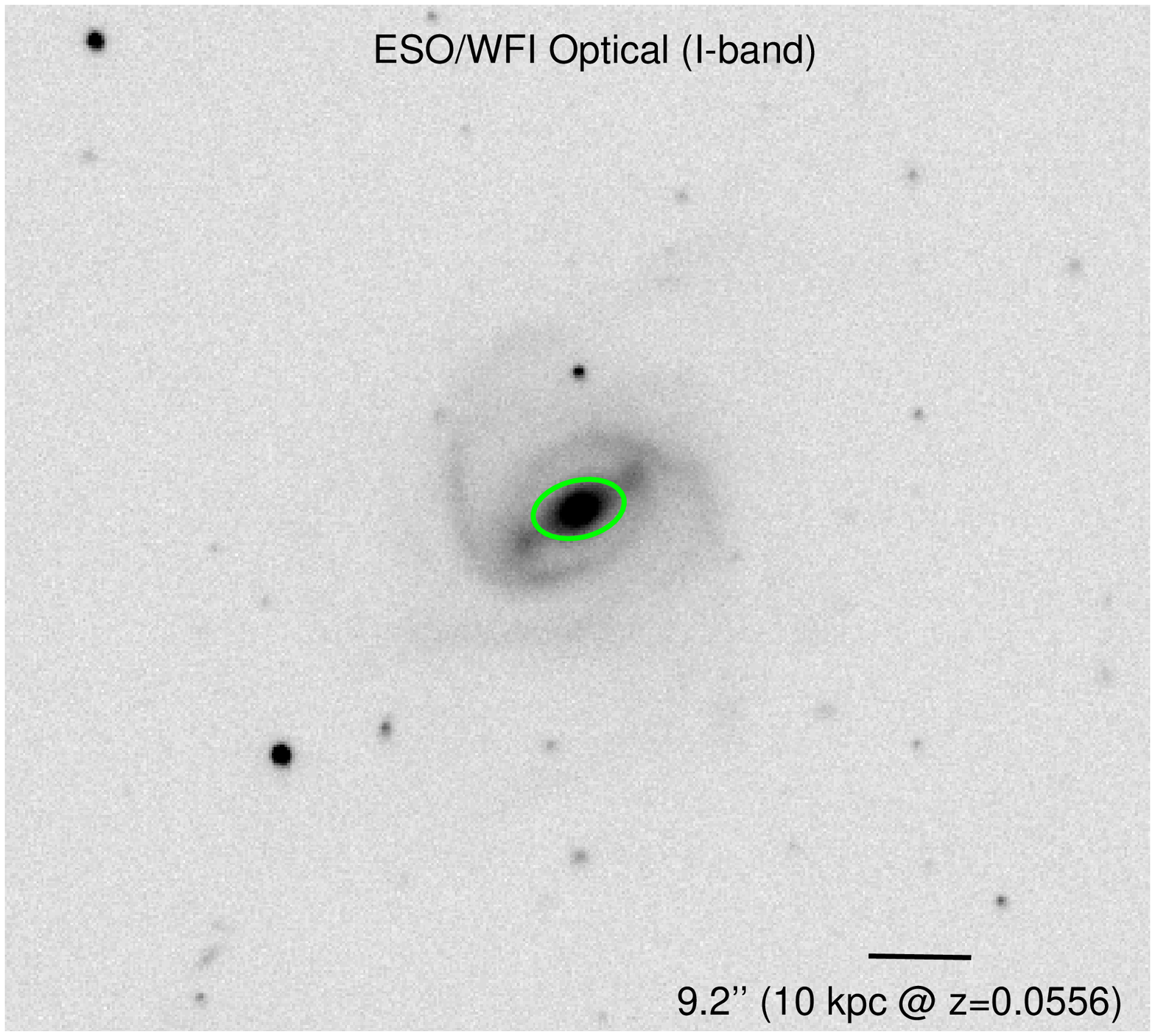}
\hskip 0.8cm \includegraphics[width=4.5cm]{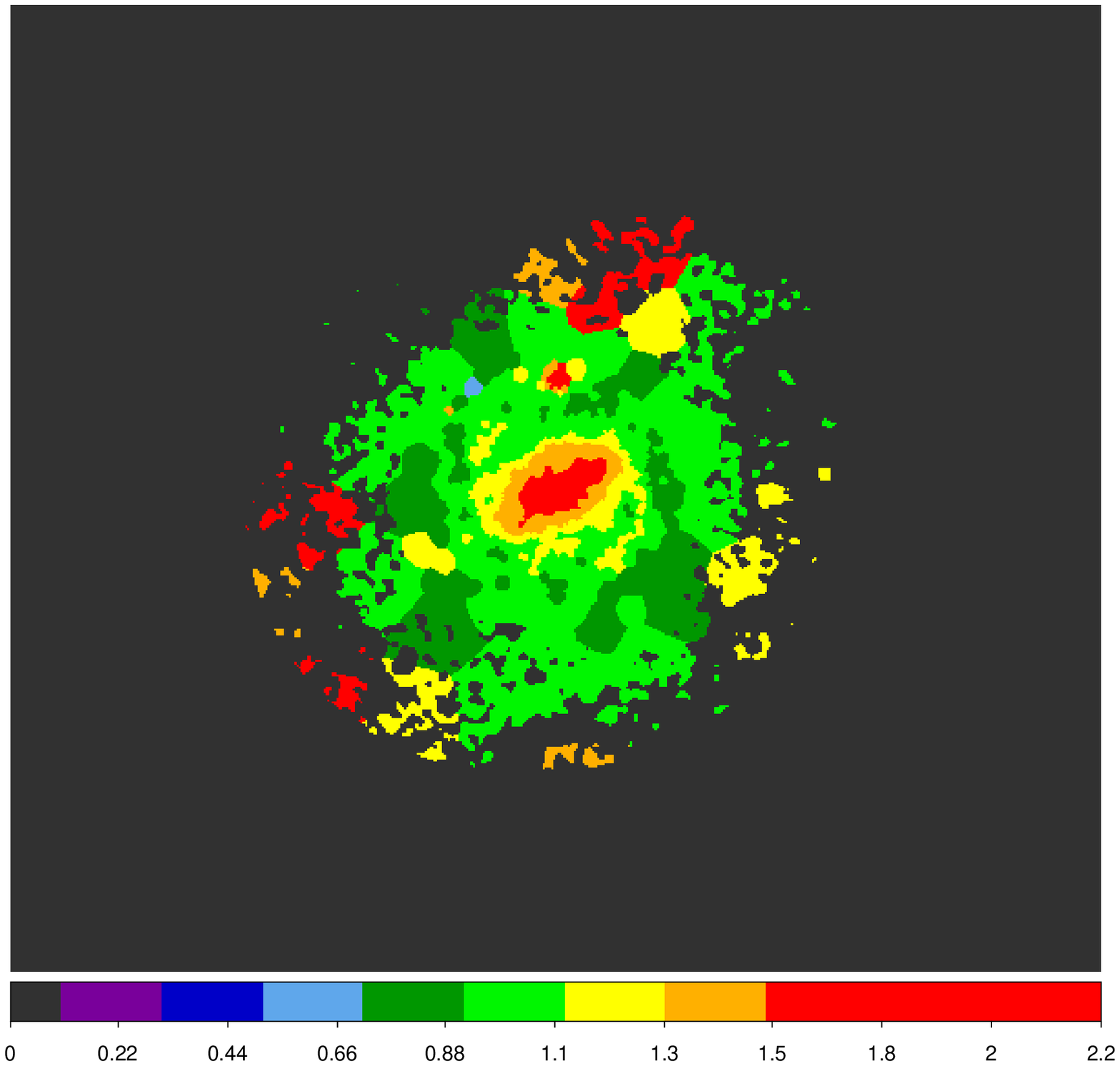}
\centering
\includegraphics[width=12cm]{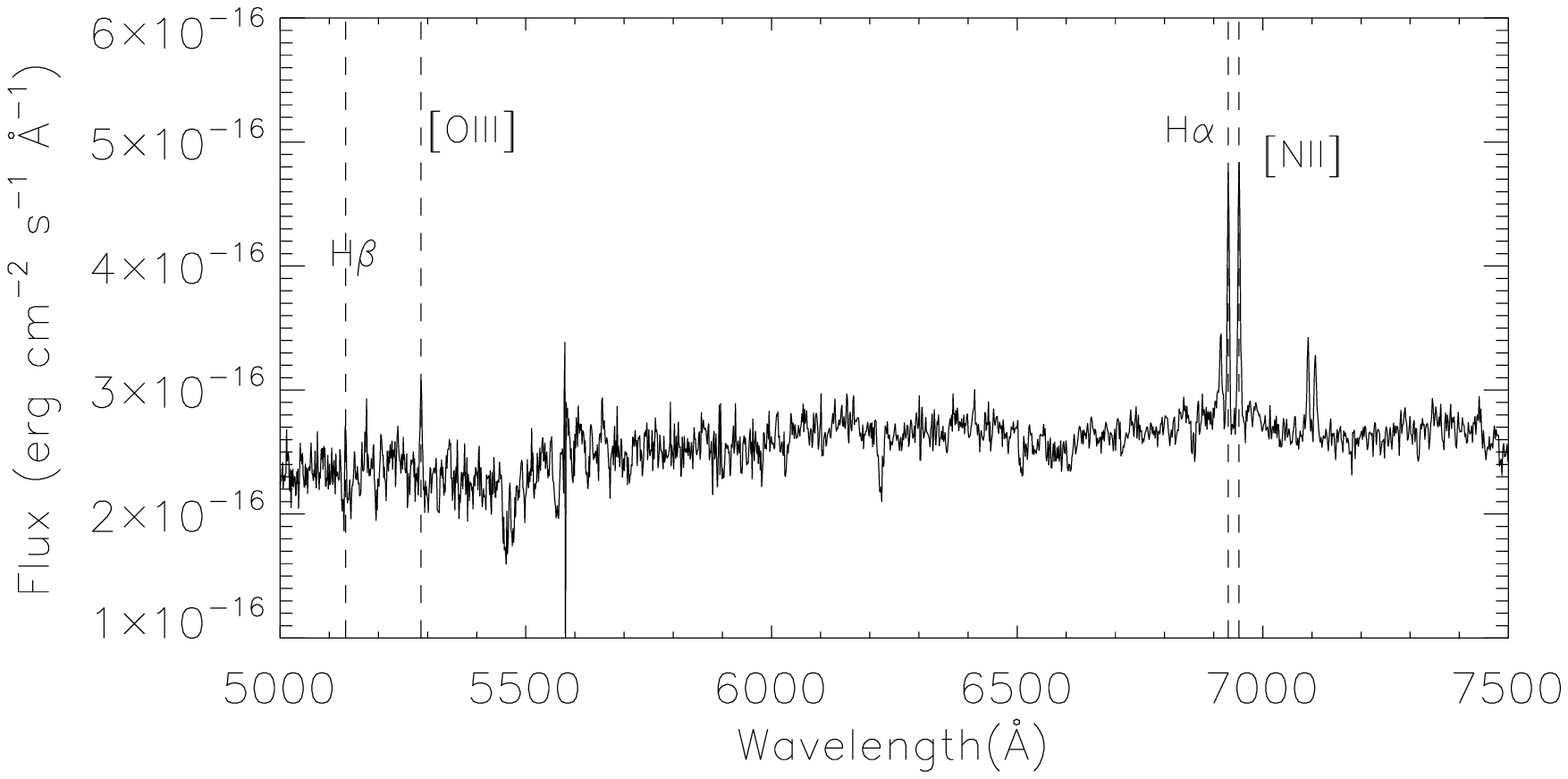}
\caption{2PIGG-n1598\_5, a massive central galaxy at z=0.0559 with a high sSFR and a point-like X-ray source.  $top~left~panel$: $Chandra$ X-ray image in the broad band (0.5-8 keV) binned by a factor of 2 for a pixel scale of 0.98$\arcsec$.  $top~middle~panel$ Optical I-band image that is matched in scale to the X-ray.  In both panels, the ellipse represents the region, as determined by 'wavdetect', that encompasses $\approx 90\% $ of the X-ray source counts that likely originates from the nucleus of this galaxy.  $top~right~panel$ Voronoi-tessellated color map \citep[see][]{Cibinel2012}.  $Bottom$ Optical spectrum from SDSS with narrow emission lines ($FWHM\sim200$ km s$^{-1}$) as labelled.}
\label{1598-5}
\end{figure*}

Before proceeding to discuss the X-ray source population, we highlight the most luminous X-ray source in our sample with $L_{0.5-8.0~{\rm keV}}=6.3\times10^{41}$ erg s$^{-1}$ with 80 X-ray counts in the broad band observation of 2PIGG-n1381 (Figure~\ref{fig:m1320}) with $Chandra$.  The optical counterpart SDSS J142808.89-023124.8 \citep{Ahn2013} is an emission-line galaxy at $z=0.0520$, as provided by SDSS, that is not included in the 2dF catalog.  However, it could be associated with 2PIGG-n1381 given its redshift ($\Delta v=60$ km s$^{-1}$ where $\Delta$$v$ is the line-of-sight velocity offset from the group center) and distance from the bright central galaxy ($\Delta r_{proj} = 57$ kpc).   The stellar mass ($M_{galaxy}=6.9\times10^{9}$ $M_{\odot}$) and specific star formation rate (sSFR = SFR/M$_{galaxy}$ = $1.7\times10^{-11}$ yr$^{-1}$) of its host galaxy, taken from the ZENS data catalog \citep{Carollo2013},  place it within the moderately star-forming population (see next section).  The optical emission line ratios are indicative of a composite nature (AGN+star forming); we infer that the X-ray emission is  predominantly due to an AGN, given that it exceeds by three orders of magnitude the typical X-ray emission from   a normal  star-forming galaxy (see details   below).  The presence of a weak broad component to the H$\alpha$ line in the SDSS spectra provides strong further evidence for an underlying AGN.  There may possibly  be also  evidence for an offset of the broad from the narrow component of H$\alpha$, that may indicate an underlying double-peaked profile \citep[e.g.,][]{Strateva2003} typically seen in low-luminosity AGNs \citep{ho2000} (or  possibly a more exotic event such as a gravitational recoil, \citep[e.g.,][]{Civano2010}).  This object exemplifies the importance of X-ray selection to cleanly identify galaxies harboring AGN of such low luminosity.

\subsection{AGN versus normal galaxies as the origin of X-ray emission}

The level of X-ray emission in ZENS galaxies is bordering on the overlap between accretion onto a black hole, stellar remnants and diffuse hot gas.  In addition to AGN, point-like X-ray emission may be attributed to the cumulative contribution of low-mass and high-mass X-ray binaries.  There is additionally room for a component of thermal emission in early-type galaxies.  The high-mass X-ray binary population is likely to contribute significantly only to those galaxies with high star formation rates.  There are many studies \citep[e.g.,][]{Ranalli2003,OSullivan2003,Lehmer2010,Boroson2011} of X-ray emission from normal galaxies (i.e., without AGN) that enable  us to determine whether the level of X-ray emission is characteristically higher than that expected for galaxies of a given luminosity, stellar mass and star formation rate.  While we are mainly concerned with nearby galaxies, such relations between X-ray emission and galaxy type appear to be universal and extend over a wide range of redshift \citep{Lehmer2007,Lehmer2008}.    

We first investigate the nature of the X-ray emission by splitting the galaxy samples into either star-forming or quiescent as classified by the presence of prominent emission lines in the optical spectra combined with NUV-NIR colors \citep{Cibinel2012}.  The diversity of galaxy spectral type within the 18 X-ZENS groups can be seen in Figure~\ref{gal_prop} where we plot the  sSFR as a function of their stellar mass $M_{galaxy}$ as done in Figure 8 of \citet[][]{Cibinel2012}.  It is worth pointing out that the ZENS sample does have incompleteness at low masses that cannot be neglected;  the lack of galaxies with $M_{galaxy}\sim10^9$ $M_{\odot}$ and sSFR between $\sim10^{-11}-10^{-12}$ yr$^{-1}$ is clearly evident.  In the figure, we indicate with an open circle those galaxies that are detected in X-rays. There is a wide spread in the distribution of sSFR at a constant stellar mass for galaxies with X-ray emission, with  X-ray sources associated with galaxies  at the extreme ends, i.e., either strongly star-forming or quiescent.  Within the limited statistics, the relative numbers of quiescent versus star-forming galaxies that emit X-rays   reflects the relative number density of the underlying galaxy population.  While the hosts span a wide range in mass from $10^9$ $M_{\odot}$ to above $10^{11}$ $M_{\odot}$, there is a preference for massive galaxies that one would expect given the well-established increase of AGN activity as a function of stellar mass \citep[e.g.,][]{Kauffmann2003,Silverman2009b,Haggard2010}.  The significance of a dependence of AGN activity on the stellar mass of their hosts is further addressed in Section~\ref{rank}.

In Figure~\ref{xtype}, the X-ray luminosity is shown as a function of  star-formation rate for galaxies classified as either highly or moderately forming stars (panel $a$), and of $B$ luminosity for the quenched population (panel $b$).  For each ZENS galaxy without an X-ray detection, we indicate an upper limit on the broad-band X-ray luminosity. For those galaxies covered by $Chandra$,  this is equivalent to our limit on a source detection of four counts at the position on the detector thus accounting for variations in the effective area as a function of off-axis angle.  The upper limits on the XMM-$Newton$ non-detections are based on a flux level at 2$\sigma$ above the background.  For reference, we provide the typical scaling relations for normal galaxies \citep{Ranalli2003, OSullivan2003} between these quantities plus an interval of $\pm1\sigma$ to indicate the typical spread.        
  
\begin{figure}
\includegraphics[angle=-90,scale=0.4]{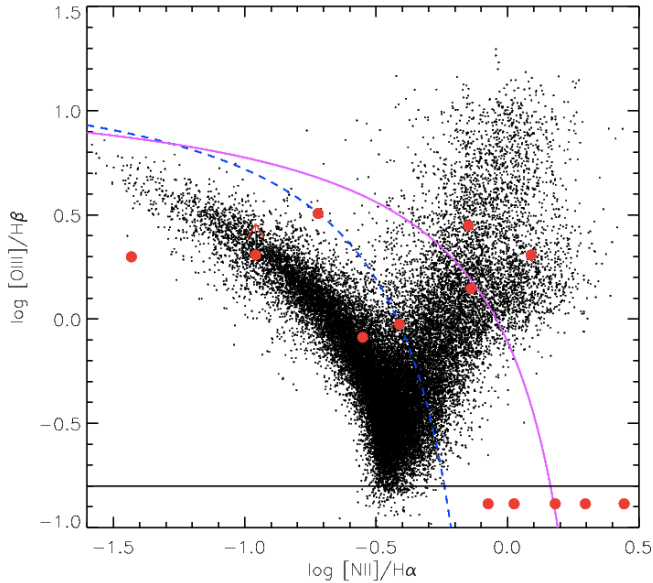}
\caption{Emission-line ratio diagram: [OIII]$\lambda$5007/H$\beta$ vs. [NII]$\lambda$6584/H$\alpha$.  Black points indicate emission-line galaxies from SDSS, and red points show the location of our X-ray ZENS sample.  The curves are the division set between star-forming galaxies and AGN as established by \citet{Kauffmann2003} (dashed) and \citet{Kewley2006} (solid).  We place galaxies in our sample with only constraints on the ratio of [NII]$\lambda$6584/H$\alpha$~at a fixed value of [OIII]$\lambda$5007/H$\beta$=-0.8}
\label{bpt}
\end{figure}

\begin{figure}
\includegraphics[angle=0,scale=0.5]{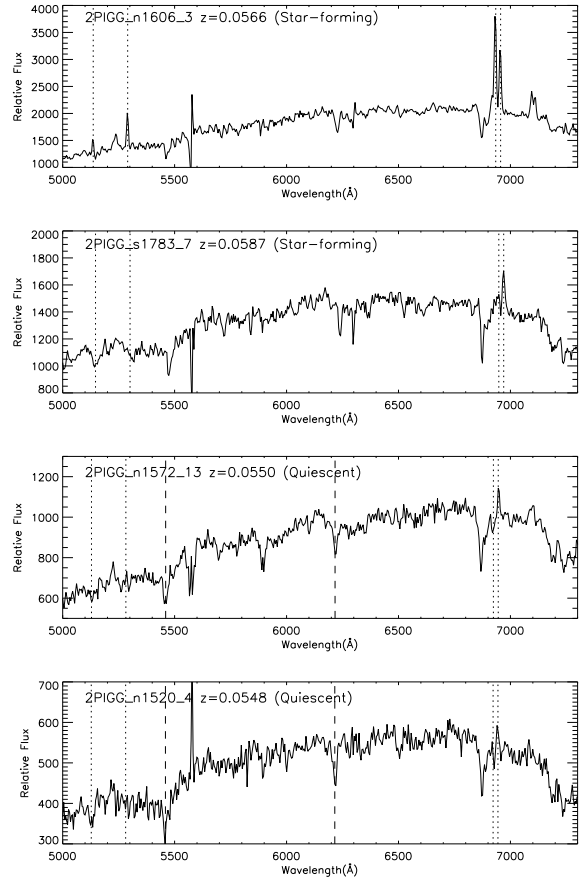}
\caption{Optical spectra from 2dFGRS of four galaxies that likely host an AGN based on their having both X-ray emission and optical line ratios indicative of such nuclear activity.  Examples are shown of galaxies classified in ZENS as either star-forming (2PIGG\_n1606\_3, 2PIGG\_s1783\_7) or quiescent (2PIGG\_n1572\_13, 2PIGG\_1520\_4).  Dotted lines mark the location of the emission lines H$\beta$, [OIII]$\lambda$5007, H$\alpha$ and [NII]$\lambda$6584.  The dashed lines in the lower two panels indicated absorption features (MgI and Na) that further confirm the redshift.}
\label{opt_spec}
\end{figure}

With respect to 111 star-forming galaxies (Fig.~\ref{xtype}$a$) falling within the $Chandra$ and XMM-$Newton$ coverage, we find that 5 out of 10 galaxies with X-ray detections are above the relation for normal, star-forming galaxies, by at least 2$\sigma$ deviation, thus likely indicative of an AGN contribution effectively boosting the X-ray luminosity above our detection limits.  From the distribution of SFRs of the X-ray non-detections, it is clear that the majority of the galaxies in ZENS have SFRs low enough that if we do detect an X-ray source, it is most probably due to an AGN.  We have detected with $XMM-Newton$ X-ray emission from three galaxies falling within the region ($\pm1\sigma$) of a normal star-forming galaxy; based on their optical emission-line properties (see below), all of these are indeed likely to have a contribution to the X-ray emission from an AGN.

We highlight the fourth such galaxy (2PIGG-n1598\_5) falling within the locus of normal star-forming galaxies.  It has the highest SFR of all ZENS galaxies observed by $Chandra$.  With a sSFR $2.6\times10^{-10}$ yr$^{-1}$ at its stellar mass of $M_{galaxy}=5.9\times10^{10}$ $M_{\odot}$, it has been identified as the central galaxy (with evidence for a strong bar) within its group.  We can make use of the high resolution imaging of $Chandra$ and quality optical WFI-ESO imaging from ZENS to determine if the X-rays are coming from the nuclear region or the overall extent of the galaxy.  In Figure~\ref{1598-5}, we show the X-ray and optical images of 2PIGG-n1598\_5.  It is clear that the X-ray emission is co-spatial with the nucleus.  The emission line ratios of [NII]/H$\alpha$ and [OIII]/$H\beta$, seen in an SDSS spectrum (Fig.~\ref{1598-5} $bottom$), place this object right on the border between AGNs and star-forming galaxies as defined by \citet{Kewley2006}, although, the [SII]/H$\alpha$ ratio might be in principle   more typical of HII regions.  We can also use the H$\alpha$ emission (detected within the SDSS fiber that covers only the very central region of the galaxy due to the fiber diameter of 3$\arcsec$) to make a more accurate assessment on the upper limit to the level of star formation in the nuclear region cospatial with the X-ray emission.  Based on a H$\alpha$ luminosity of $8.4\times10^{39}$ erg s$^{-1}$, we estimate the star formation rate of the central region of the galaxy to be around 0.07 M$_{\odot}$ yr$^{-1}$ \citep{Kennicutt1998}, substantially lower than that of the entire galaxy and not likely to produce the amount of X-rays observed by $Chandra$.  Therefore, we conclude that a low-luminosity AGN with $L_X=8\times10^{40}$ erg s$^{-1}$  is powering the X-ray emission seen in 2PIGG-n1598\_5.

\begin{figure*}
\epsscale{1.0}
\plotone{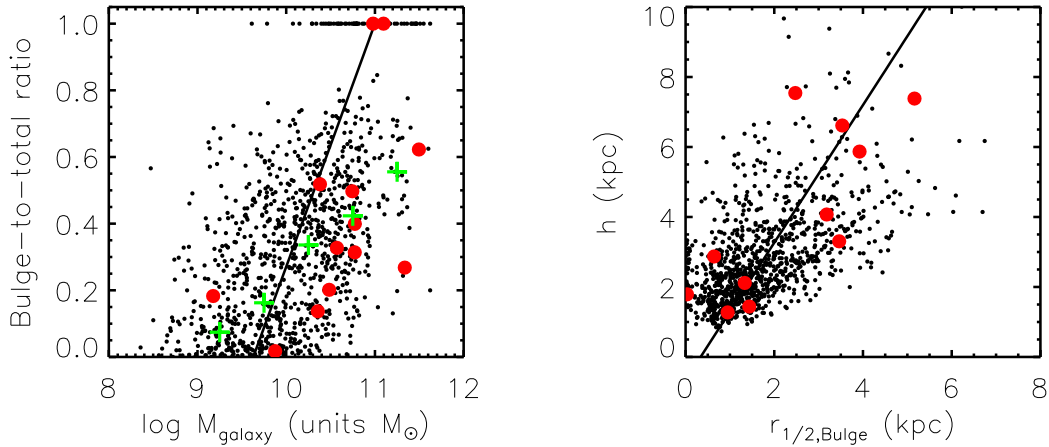}
\caption{$Right:$ Bulge-to-total ratio versus galaxy stellar mass of ZENS galaxies (black points; elliptical galaxies are placed at B/T=1).  Median B/T ratios in bins of stellar mass are indicated by the green crosses.  $Left:$ For galaxies with both a bulge and a disk component, we plot  the disk scale length $h$ versus  the half-light radius of the bulge $r_{1/2,Bulge}$, both in units of kpc.  Both panels show with a solid line the best-fit linear relation for the whole galaxy population, with  the AGNs  marked as larger red circles.}
\label{morph2}
\end{figure*}

Out of the 66 quiescent galaxies observed in X-rays, there are 12 detections with 10 of them falling with a region of the $L_X-L_B$ plane spanned by normal galaxies (see Figure~\ref{xtype}$b$).  These galaxies need not necessarily have an AGN to explain their level of X-ray emission as was primarily the case for the star-forming population.  As done above, we can measure the spatial extent of the X-ray emission, based on $Chandra$ imaging and determine whether it is extended beyond that expected based on the point spread function (PSF) at the given position on the detector since the size of the PSF is a strong function of the off-axis angle.  Based on the three objects having $Chandra$ imaging, we find that two (2PIGG\_n1347\_3 and 2PIGG\_n1671\_4) are clearly extended while 2PIGG\_s1752\_4 has a spatial count distribution similar to that expected due to an unresolved point source.  Therefore, we are able to classify the latter as an AGN.
                                   
As an independent  test to further solidify the presence of an AGN, we make use of the optical emission line properties of ZENS galaxies using the 2dFGRS spectra.  We specifically measure the emission line ratios [NII]$\lambda$6584/H$\alpha$, and [OIII]$\lambda5007$/H$\beta$ if both lines are detected.  In Figure~\ref{bpt}, we show the distribution of their ratios as compared to emission-line galaxies from SDSS.  For many galaxies in our sample, the H$\alpha$ and [NII] emission lines are clearly present while the H$\beta$ and [OIII] lines are very faint, primarily for the quiescent galaxy population.  In Figure~\ref{opt_spec}, we provide examples of the 2dFGRS spectra of galaxies, classified in ZENS as either star-forming or quiescent, that have X-ray emission placing them within the locus of normal galaxies (Figure~\ref{xtype}), and optical line ratios typical of an AGN.  A limitation is that we cannot separate Seyferts and LINERS, where the latter have been heavily debated in the literature as to whether their line ratios are indicative of AGN photoionization or that due to post-AGB stars \citep{Sarzi2010,Yan2012}.  A reasonable working hypothesis, mainly pertaining to quenched galaxies, is that a high [NII]/H$\alpha$ ratio is an indication of an AGN, irrespective of the  H$\beta$ and [OIII] line strengths.  In Figure~\ref{bpt}, we find that 10 of the 13 galaxies for which we could measure a [NII]/H$\alpha$ emission-line ratio are unlikely to be typical star-forming galaxies, given their location relative to the demarcation line of \citet{Kauffmann2003}.  The optical spectra of the remaining 9 galaxies with X-ray detections have either very weak or non-existent H$\alpha$ or [NII]$\lambda$6584 emission.

Summarizing, our final decision on whether X-ray emission is attributed to an AGN is based on three lines of evidence.  First and foremost, if there is X-ray emission substantially higher than that expected from normal star-forming or quiescent galaxies.  Second, if the X-ray emission is not diffuse as determined with $Chandra$ resolution (otherwise, if diffuse, it is attributed to thermal emission).  Third, we consider the optical emission line ratios to aid in the discrimination between photoionization indicative of an AGN and UV emission from young stars.  Taking these together, we find that the 16 of 22 X-ray sources are associated with emission from an AGN.  Those that are not AGN are mostly quiescent galaxies with either noticeable thermal emission or unresolved stellar remnants.  Our detection rate for AGNs  is consistent with the steeply rising faint end of the local AGN X-ray luminosity function \citep{Sazonov2004,Ueda2011}, for which our sample spans $L_X\sim0.4 - 6.6\times 10^{41}$ erg s$^{-1}$, and with the area coverage of the X-ZENS  observations.

It is important to recognize that the X-ray emission may be of a composite nature with AGN, stellar or thermal processes all making some contribution to the total X-ray emission.  This may provide a boost in the number of X-ray detections as expected from normal galaxies, complicate the selection of AGN, and hamper the determination of a pure AGN luminosity used to determine a bolometric quantity (Section~\ref{host-mass}).  With the limited sample in hand, there is little that can be done to quantify sufficiently such effects.  A larger sample will allow us to determine at what luminosities does such dilution of the X-ray emission become problematic.  There may even be X-ray data in the archive of local galaxies where this can be addressed more effectively than with the ZENS sample.

Finally note that we  detect X-rays from two (satellite) galaxies with low stellar masses $M_{galaxy}\sim10^9$ $M_{\odot}$ (see Figure~\ref{gal_prop}). If the X-rays are attributed to black hole accretion, the black holes are likely to be of low mass $M_{BH}\sim10^6$ $M_{\odot}$ with Eddington rates below $10^{-2}$ (see Figure~\ref{lbol-mass}, obtained assuming  a typical bulge-to-disk ratio  of $\sim 0.5$  and  local scaling relations; \citealt{Haering2004}). Recently, \citet{Schramm2013} have reported on the discovery of three low-mass galaxies with $M_{galaxy}<3\times10^9$ $M_{\odot}$  in the Chandra Deep Field   South survey that emit X-rays which  are likely attributed to AGN activity (with $M_{BH}\sim~a~few~\times10^5$ $M_{\odot}$).  Such low black holes  mass and accretion rate regimes are still largely unexplored \citep[see][for a recent effort using optical selection]{Reines2013}; this shows the potential for opening  them to detailed studies of X-ZENS-like surveys extended however to larger group samples.

\section{The AGN  content of  the X-ZENS groups}

While the main focus of this first X-ZENS paper is to describe our current X-ray observations and point-source detections, we briefly  carry out a preliminary analysis of the demographics of our AGN sample with respect to the larger ZENS database and in comparison with the field. Specifically,  we discuss below   whether AGNs in galaxy groups show any preference for any given type of galaxy hosts and environments,  i.e.,  galaxies of different bulge-to-total ratio, central or satellite rank within the halos, and   at small or large halo-centric distances.

\subsection{Structural properties  of  ZENS galaxies hosting AGNs}

We first determine whether galaxies in groups with X-ray luminosities above our detection  threshold  are preferentially hosted by  any particular type of galaxy host.  The left panel of Figure \ref{morph2} shows, as a function of galaxy stellar mass,  the bulge-to-total ratio $B/T$, when available,  as derived from double-component fits to the bulge and disk light distributions. These were described with, respectively, a general S\'ersic profile \citep{Sersic1968} and an exponential profile \citep[see][]{Cibinel2013}. The whole ZENS galaxy sample is reported with black points, and in red are highlighted galaxies with an X-ray detected AGN\footnote{From this analysis we exclude SDSSJ142808.89-023124.8, which is not included in the 2DF sample, to avoid introducing biases in our assessment.}.    Galaxies with a purely elliptical morphology are placed at a bulge-to-total value of 1.  For reference, we show the median ratio B/T for galaxies with double component fits in bins of stellar mass ($10^{9}<M<10^{11.5} M_{\odot}$; $\Delta log~M=0.5~M_{\odot}$).  For galaxies which have both a bulge and a disk component, the right panel of the same figure shows the disk scale-length $h$ plotted versus the half-light radius of the bulge, $r_{1/2,Bulge}$.  Interestingly, the AGN hosts appear to be on average slightly `under-bulged' relative to the galaxy population at similar stellar masses (i.e., the AGN hosts systematically lie in the bottom half of the $B/T$ versus galaxy stellar mass relation); on the other hand, the bulges of AGN hosts appear to be similarly  `embedded'  within their surrounding disks than the global galaxy population at similar mass scales (i.e., they are distributed evenly around the best fit to the $h$ vs $r_{1/2,Bulge}$ relation. Possible explanations are that,  at a given mass, the hosts of AGNs have either relatively brighter/denser disks or relatively fainter/diffuse bulges than their non-active relatives.

\begin{figure*}
\centering
\includegraphics[angle=0,scale=0.6]{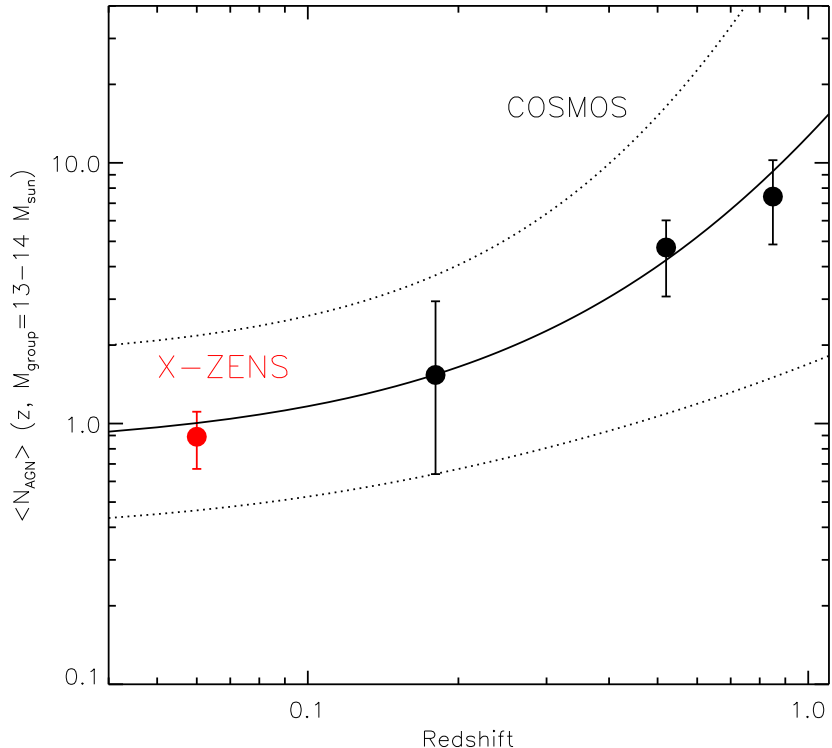}
\caption{Mean number of AGNs per galaxy group (i.e., halo occupation) in the  $13< log~M_{group}<14$ mass range as a function of redshift.  The X-ZENS data point is shown in red; in black are  the measurements from the COSMOS study of  \citet{Allevato2012}, scaled to match the X-ray luminosities of our sample. The solid line indicates a best-fit relation (dotted lines at $\pm1\sigma$) with an evolution rate $\propto(1+z)^{3.99}$, i.e., similar to the evolution of the global AGN population.}
\label{hod}
\end{figure*}

\subsection{Halo occupation distribution}

We measure the halo occupation distribution (HOD) of AGNs in the X-ZENS groups.  With 16 AGNs distributed in 18 groups,   the mean number of AGNs  with $L_X>10^{40}$ erg s$^{-1}$  per group is $0.89\pm0.22$ (with the error based on Poisson number counts).  We compare the number of AGNs detected in our groups to those expected based on the X-ray luminosity function (XLF) of the global AGN population \citep{Ueda2011}, to establish whether an over-density of AGNs is seen within group-sized potentials.  We restrict the analysis to AGNs found in the $Chandra$  observations of the 12 groups, since we   implemented an identical X-ray count threshold as in the observations of the Bootes survey \citep{Kenter2005}, and can thus make use of the well-established sky area curve as a function of X-ray flux sensitivity. We find a surface density of 9.2 AGNs per square degree considering X-ray detections within 8.5\arcmin~of the aim point.  We then integrate the XLF over the narrow redshift range  $0.05<z<0.06$ and over X-ray luminosities between $10^{40.3}-10^{46}$ erg s$^{-1}$.  Based on the XLF, we would expect to detect 3.1 AGNs per square degree with redshifts and luminosities within these ranges.  The  enhancement of AGNs of a factor $\approx3$ relative to the XLF-based estimates  is in good agreement with the typical galaxy over-densities of the X-ZENS  groups, as shown in Figure~\ref{group_prop}, indicating no extra  AGN activity in groups relative to the global population.  

Alternatively, we can compare the fraction of galaxies hosting an AGN to published studies at similar redshifts and over a range of environmental conditions.  For massive clusters, \citet{Martini2006} find that the AGN fraction of galaxies with $M_R<-20$ is $\sim5\pm1.5\%$.  Using a similar magnitude selection and scaling our results to match their coverage of the AGN XLF, we measure an AGN fraction of $8.2\pm2.1\%$.  While there might be slight evidence for an enhancement of the AGN fraction in groups relative to the more massive clusters (fully compatible with a similar analysis presented in \citet{Arnold2009}), a larger X-ray selected AGN sample in ZENS groups is clearly needed to statistically substantiate such a claim.  For a comparison to less dense environments, we find that the AGN fraction of galaxies in ZENS groups ($\sim2\%$; 1 out of 54 galaxies with $-21<M_i<-20$) is compatible with estimates of the AGN  fraction of galaxies in the field \citep{Haggard2010};  such a narrow selection of absolute magnitude is required to compare to their results (i.e., sample 2 in their study).

We then determine how the AGN HOD of X-ZENS groups compares to higher redshift measurements.  \citet{Allevato2012} provide the AGN distribution in X-ray selected groups in COSMOS up to $z\sim1$. Despite the different    selection criteria for the COSMOS and X-ZENS samples, the halo mass ranges are very similar thus allowing us to make such a comparison and look for evolutionary trends.  For this exercise, we assume that the shape of the 0.5-8 keV XLF does not change with redshift, and that  the faint-end slope is a strict power law below $L_X\sim10^{42}$ erg s$^{-1}$.   The first assumption is supported by observational evidence up to $z\sim1$ \citep[e.g.,][]{Hasinger2005,Silverman2008, Aird2010}, where the evolution has been shown to be essentially a global shift in luminosity (i.e., pure luminosity evolution).  The second assumption is based on an extrapolation of the known XLF.  We  use the XLF of \citet{Ueda2011}  to derive a scale factor which accounts for the difference in flux (or luminosity) sensitivity between the COSMOS and ZENS X-ray data sets.  A scaling of the AGN HOD of 12.8 is required to match the two samples down to the luminosity limit of our X-ZENS data.

In Figure~\ref{hod}, we plot the mean number of AGNs per halo as a function of redshift and fit the data with a function having terms for the normalization and redshift evolution.  We find the following relation with best-fit parameters and $1\sigma$ errors:

\begin{equation}
<N_{AGN}>=0.80_{-0.43}^{+0.89}\times(1+z)^{3.99\pm2.69}
\end{equation}

\noindent Even considering the uncertainties, we find that there is very good agreement between the two samples as indicated by the redshift evolution of the AGN HOD, shown by the solid curve (Fig.~\ref{hod}).  Furthemore, the rate of evolution is practically identical to the global X-ray luminosity function of \citet{Ueda2003} which has evolution parameterized as $(1+z)^{4.2}$, which is also well reproduced by other determinations for the AGN population \citep[][]{Silverman2008, Ebrero2009,Aird2010}.  We conclude that the rate of decline with redshift of AGN activity in galaxy groups is similar to that of the global AGN population. The origin of this similarity may rest on  either a predominance of AGN activity in halos at the  $M_{group}\sim10^{13-14} M_\odot$ mass scale, as supported by clustering analyses of AGNs (e.g.,  Porciani \& Norberg 2006), or on the independence of AGN evolution on halo mass.

\begin{figure}
\includegraphics[angle=90,scale=0.35]{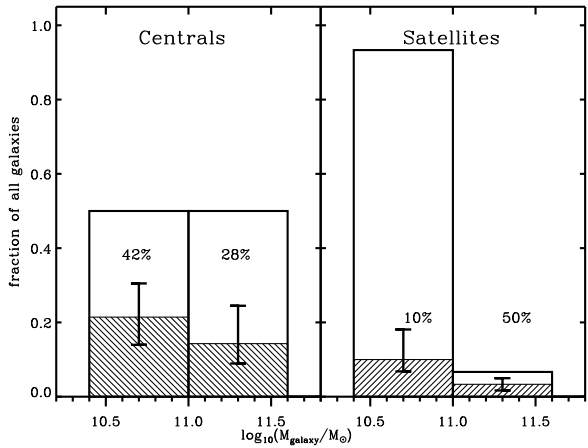}
\caption{Fraction of all galaxies split into two mass bins and presented as central (left) and satellite (right) galaxies separately (open histogram).  A mass range of $10^{10.4}M_{\odot}<M_{galaxy}<10^{11.6} M_{\odot}$ is chosen where both the centrals and satellite populations are well-represented in ZENS.  The fraction of galaxies in each bin that host an AGN is indicated by hatched histograms and associated error bars. All histograms are normalized to the total number of galaxies (within that rank bin) in the mass range $10^{10.4}M_{\odot}<M_{galaxy}<10^{11.6} M_{\odot}$.  The AGN fractions within each mass and rank bin are given as percentages.}
\label{cen-sat}
\end{figure}

\begin{figure}
\includegraphics[angle=90,scale=0.35]{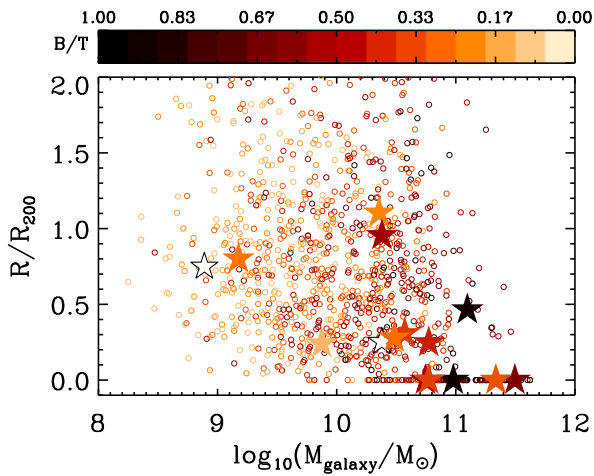}
\caption{ Halo-centric distance $R$ of all ZENS galaxies, irrespective of having X-ray coverage, in units of ${\hat R}_{200}$ (small open circles), versus galaxy stellar mass.  Galaxies hosting X-ray detected AGNs are plotted as large stars.  The color coding is illustrative of the bulge-to-total ratio (B/T).  Open black stars have no discernible measurement of B/T.} 
\label{distance}
\end{figure}

\subsection{AGN hosts: Centrals, inner satellites  or outer satellites?}

\label{rank}

Next, we  investigate whether AGN activity is typical of central or satellite galaxies within groups, and whether it is enhanced in the cores or outskirts of the typical $\sim10^{13} M_\odot$ galaxy groups that we probe in X-ZENS.  There is evidence in a number of other studies \citep[e.g.,][]{Ruderman2005,Martini2007,Martel2007,Fassbender2012}  that AGN   show a preference for the inner regions of   massive clusters, at halo masses larger than the X-ZENS values, and that AGNs may possibly be associated with the brightest cluster members.  There are however other studies  which report opposite results, i.e., even possible AGN excesses in the outskirts of galaxy clusters \citep[e.g.,][]{Gilmour2009, Pentericci2013}.  At the lower halo mass scales of our sample,   cosmological simulations  predict that AGN activity should be  more closely tied to central galaxies as opposed to satellites \citep{Richardson2013}.

\begin{figure}
\epsscale{0.9}
\plotone{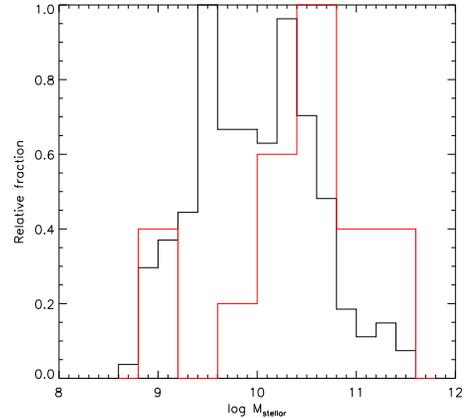}
\caption{Stellar mass distribution of galaxies with X-ray observations (black histogram) and those that host AGN (red histogram).  Both distributions have a peak set to unity.  AGNs are clearly preferential to more massive galaxies.}
\label{mass-distr}
\end{figure}

We  quantify  the fraction of galaxies, split as central and satellites, that host an AGN over the total mass range $10.4\ge M_{galaxy} \le 11.6$. We find that 5 of 14 central galaxies host an AGN ($0.36_{-0.11}^{+0.14}$), and only 4 out of  30 satellite galaxies  host  an AGN ($0.13_{-0.04}^{+0.09}$).  Thus, although there is a   hint for an enhancement by a factor of $\sim3$ for AGNs in centrals relative to satellites, as predicted by the simulations, this is detected in our data at a low statistical level of significance. A clearer effect is however seen when we investigate the relative frequency of AGNs in centrals and satellites in narrower stellar mass bins, as listed  in Table~\ref{table:rank}. In particular, at galaxy mass scales $<10^{11} M_\odot$,  there is a clear effect for centrals hosting AGNs about four times more frequently than satellites of similar mass., as shown in Figure~\ref{cen-sat}.    

 In Figure~\ref{distance}  we furthermore explore whether there is any differential effect with halo-centric distance in the frequency of AGNs in satellites. The figure shows  halo-centric distance $R$ in units of the group characteristic radius ${\hat R}_{200}$ \citep[see][for an explicit definition of this parameter]{Carollo2013} as a function of stellar mass.  The majority of the detected AGNs (11/15), as shown by the star symbols, are within $R/{\hat R}_{200}<0.5$. This may however be possibly explained by the   number of galaxy and galaxy mass  variations with halo-centric distance (including differences between centrals and satellites). Indeed,  a 2-D K-S test returns no significant  difference between the radial distribution  of AGNs and galaxies  at stellar masses above $10^{10.4} M_\odot$.  Also, a
  2-D KS test on the stellar mass distribution of galaxies targeted in X-rays and those hosting AGNs shows that these distributions are dissimilar at the 3.3$\sigma$ level, as shown in Figure~\ref{mass-distr}: AGNs are indeed more prevalent in massive galaxies ($log~M_{galaxy}\gtrsim10.3$), as independently seen in other studies.
  
\subsection{Black hole masses and Eddington ratios}
\label{host-mass}

\begin{figure}
\includegraphics[angle=90,scale=0.45]{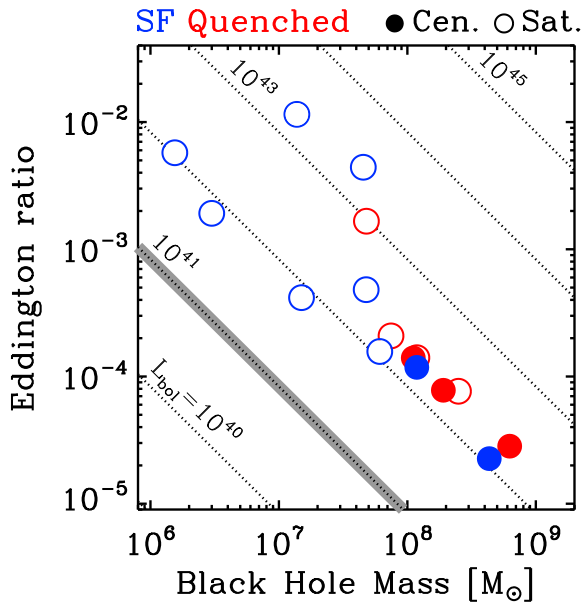}
\caption{Eddington ratio versus black hole mass as derived from the stellar mass of the host galaxies and an application of known scaling relation \citep{Haering2004}.  Colors represent whether a galaxy has been quenched (red) or not (blue) with a definition set as having an sSFR above or below $10^{-11}$ yr$^{-1}$. Central and satellite galaxies are indicated by filled and open symbols.  Lines of constant bolometric luminosity are provided including the effective limit of our survey at L$_{bol}=10^{40}$ erg s$^{-1}$ shaded in grey.}
\label{lbol-mass}
\end{figure}

Finally we assess in what part of parameter space our AGN fall in terms of their likely black hole masses and Eddington rates.  To do so, we use the local scaling relation between black hole mass and the stellar mass of its host galaxy \citep{Haering2004}.  Bolometric luminosities are derived by applying correction factor of 20 to the broad-band X-ray luminosity as applicable to X-ray selected AGNs in COSMOS \citep{Lusso2012}, with a caveat that it is an area of debate whether  this factor is appropriate for low luminosity AGNs.    In Figure~\ref{lbol-mass}, we find that the majority of our sample falls well below the Eddington rate with $L/L_{Edd}\sim10^{-4}$.  This is  2-3 orders of magnitude below more luminous quasars such as those in SDSS \citep[e.g.,][]{Shen2011,Kelly2013} and AGNs in COSMOS \citep{Trump2009,Lusso2012}, i.e., much lower than these latter species, even assuming a large correction factor to the normalization value used above.  Disentangling in Figure~\ref{lbol-mass}  quenched galaxies  (with sSFR$<10^{-11}$ yr$^{-1}$) from star forming galaxies, and centrals from satellites,  it is clear that  the X-ZENS AGNs with the lowest Eddington ratios ($<10^{-4}$) are hosted at the high mass end of the galaxy population.  Already our small samples shows that such `starved' AGNs can occur both in central and satellite galaxies (albeit with a possible preference for centrals), and in quenched as well as star forming hosts (albeit with a preference for quenched systems).
 
\section{Summary}

We have  presented the  X-ray observations of 18  of our $z\sim0.05$ ZENS  galaxy groups. The observations were taken using both $Chandra$ and XMM-$Newton$. This paper has focused on the data acquisition, analysis and point-source catalog.  With $Chandra$ exposures of 10 ksec each, we reach a depth sufficient to detect AGNs down to $L_{0.5-8~{\rm keV}}\gtrsim3\times10^{40}$ erg s$^{-1}$ at $z\sim0.05$.  Our XMM-$Newton$ data reaches comparable depths due to the requirement to significantly detect diffuse emission from the DIM as presented in a companion X-ZENS paper (Miniati et al. in preparation).  In total, we detect X-ray emission from 22 out of 177 galaxies targeted in X-rays.

We distinguish the origin of the X-ray emission as due to either an AGN, or thermal emission or from stellar remnants.  We find that X-ray emission seen in strongly star-forming galaxies, unresolved in all cases, is likely due to an AGN in 9 out of 10 cases,  even in galaxies of low mass that may harbor low mass black holes.  X-ray emission in quiescent galaxies is seen to have a significant contribution from hot diffuse gas and/or a cumulative signal from stellar remnants (i.e., low mass X-ray binaries) based on the X-ray emission at expected levels comparable to normal (non-AGN) galaxies, and in some cases, spatially extended as compared to that expected due to PSF variations across the focal plane.  Of the quiescent galaxies, we associate 7 out of 12 galaxies with AGN activity.  In total, we find that 16 of the 22 X-ray sources are likely due to accretion onto a SMBH, down to very low Eddington rates of $\sim10^{-4}$.  

While our primary aim here is to provide details on our X-ray program, we have began to address some of the scientific questions on AGN activity in groups at this mass scale.  We have measured the halo occupation density of AGNs, and found  a lower occupation fraction relative  to groups of similar mass up to $z\sim1$.  The observed  decline with decreasing redshift is entirely consistent with the known evolution of the global AGN population, and suggests  either that AGNs do inhabit preferentially such intermediate group-sized halos, or that the growth of black holes in groups at this mass scale proceeds at a similar rate than in other environments.  

As also seen in other studies, AGNs tend to be hosted by massive galaxies.  At a given galaxy mass, the galaxies which host an AGN may have either relatively brighter, and thus possibly denser  disks, or relatively  fainter, and thus possibly more diffuse bulges, than  galaxies which do not host an AGN. At galaxy masses $<10^{11} M_\odot$,   AGNs appear four times more often in   central than in satellite galaxies   of similar mass, an effect which explains why  AGNs are preferentially found in the cores of groups, without any  detectable trend  in the frequency of AGNs  in satellite galaxies at different halo-centric distances.  

X-ZENS provides a low-redshift benchmark for comparisons with X-ray surveys of groups at higher redshifts, and a low-mass benchmark for comparisons with X-ray surveys of massive clusters -- two of the main scientific motivations  of our program.

 \acknowledgments

This work was supported by World Premier International Research Center Initiative (WPI Initiative), MEXT, Japan.  AC acknowledges financial support from the Swiss National Science Foundation (SNF).  KS gratefully acknowledges support from Swiss National Science Foundation Grant PP00P2\_138979/1
 
\bibliographystyle{apj}
\bibliography{myrefs}

\begin{thebibliography}{81}
\expandafter\ifx\csname natexlab\endcsname\relax\def\natexlab#1{#1}\fi

\bibitem[{Ahn} et~al.(2013){Ahn}, {Alexandroff}, {Allende Prieto}
  et~al.]{Ahn2013}
{Ahn} C.~P., {Alexandroff} R., {Allende Prieto} C., et~al., 2013, ArXiv
  e-prints

\bibitem[{Aird} et~al.(2010){Aird}, {Nandra}, {Laird} et~al.]{Aird2010}
{Aird} J., {Nandra} K., {Laird} E.~S., et~al., 2010, \mnras, 401, 2531

\bibitem[{Allevato} et~al.(2012){Allevato}, {Finoguenov}, {Hasinger}
  et~al.]{Allevato2012}
{Allevato} V., {Finoguenov} A., {Hasinger} G., et~al., 2012, \apj, 758, 47

\bibitem[{Arnold} et~al.(2009){Arnold}, {Martini}, {Mulchaey}, {Berti} \&
  {Jeltema}]{Arnold2009}
{Arnold} T.~J., {Martini} P., {Mulchaey} J.~S., {Berti} A., {Jeltema} T.~E.,
  2009, \apj, 707, 1691

\bibitem[{Barnes}(1990)]{Barnes1990}
{Barnes} J.~E., 1990, \nat, 344, 379

\bibitem[{Bielby} et~al.(2010){Bielby}, {Finoguenov}, {Tanaka}
  et~al.]{Bielby2010}
{Bielby} R.~M., {Finoguenov} A., {Tanaka} M., et~al., 2010, \aap, 523, A66

\bibitem[{Boroson} et~al.(2011){Boroson}, {Kim} \& {Fabbiano}]{Boroson2011}
{Boroson} B., {Kim} D.-W., {Fabbiano} G., 2011, \apj, 729, 12

\bibitem[{Carollo} et~al.(2013){Carollo}, {Cibinel}, {Lilly}
  et~al.]{Carollo2013}
{Carollo} C.~M., {Cibinel} A., {Lilly} S.~J., et~al., 2013, \apj, 776, 71

\bibitem[{Cibinel} et~al.(2012){Cibinel}, {Carollo}, {Lilly}
  et~al.]{Cibinel2012}
{Cibinel} A., {Carollo} C.~M., {Lilly} S.~J., et~al., 2012, ArXiv e-prints

\bibitem[{Cibinel} et~al.(2013){Cibinel}, {Carollo}, {Lilly}
  et~al.]{Cibinel2013}
{Cibinel} A., {Carollo} C.~M., {Lilly} S.~J., et~al., 2013, \apj, 776, 72

\bibitem[{Civano} et~al.(2010){Civano}, {Elvis}, {Lanzuisi} et~al.]{Civano2010}
{Civano} F., {Elvis} M., {Lanzuisi} G., et~al., 2010, \apj, 717, 209

\bibitem[{Colless} et~al.(2001){Colless}, {Dalton}, {Maddox}
  et~al.]{Colless2001}
{Colless} M., {Dalton} G., {Maddox} S., et~al., 2001, \mnras, 328, 1039

\bibitem[{Colless} et~al.(2003){Colless}, {Peterson}, {Jackson}
  et~al.]{Colless2003}
{Colless} M., {Peterson} B.~A., {Jackson} C., et~al., 2003, preprint
  (astro-ph/0306581)

\bibitem[{Dickey} \& {Lockman}(1990)]{Dickey1990}
{Dickey} J.~M., {Lockman} F.~J., 1990, \araa, 28, 215

\bibitem[{Ebrero} et~al.(2009){Ebrero}, {Carrera}, {Page} et~al.]{Ebrero2009}
{Ebrero} J., {Carrera} F.~J., {Page} M.~J., et~al., 2009, \aap, 493, 55

\bibitem[{Eke} et~al.(2004{\natexlab{a}}){Eke}, {Baugh}, {Cole}
  et~al.]{Eke2004a}
{Eke} V.~R., {Baugh} C.~M., {Cole} S., et~al., 2004{\natexlab{a}}, \mnras, 348,
  866

\bibitem[{Eke} et~al.(2004{\natexlab{b}}){Eke}, {Frenk}, {Baugh}
  et~al.]{Eke2004b}
{Eke} V.~R., {Frenk} C.~S., {Baugh} C.~M., et~al., 2004{\natexlab{b}}, \mnras,
  355, 769

\bibitem[{Ellison} et~al.(2011){Ellison}, {Patton}, {Mendel} \&
  {Scudder}]{Ellison2011}
{Ellison} S.~L., {Patton} D.~R., {Mendel} J.~T., {Scudder} J.~M., 2011, \mnras,
  418, 2043

\bibitem[{Fassbender} et~al.(2012){Fassbender}, {{\v S}uhada} \&
  {Nastasi}]{Fassbender2012}
{Fassbender} R., {{\v S}uhada} R., {Nastasi} A., 2012, Advances in Astronomy,
  2012

\bibitem[{Ferrarese} \& {Merritt}(2000)]{Ferrarese2000}
{Ferrarese} L., {Merritt} D., 2000, \apjl, 539, L9

\bibitem[{Finoguenov} et~al.(2007){Finoguenov}, {Guzzo}, {Hasinger}
  et~al.]{Finoguenov2007}
{Finoguenov} A., {Guzzo} L., {Hasinger} G., et~al., 2007, \apjs, 172, 182

\bibitem[{Gebhardt} et~al.(2000){Gebhardt}, {Bender}, {Bower}
  et~al.]{Gebhardt2000}
{Gebhardt} K., {Bender} R., {Bower} G., et~al., 2000, \apjl, 539, L13

\bibitem[{Georgakakis} et~al.(2008){Georgakakis}, {Gerke}, {Nandra}
  et~al.]{georgakakis2008}
{Georgakakis} A., {Gerke} B.~F., {Nandra} K., et~al., 2008, \mnras, 391, 183

\bibitem[{Giavalisco} et~al.(2004){Giavalisco}, {Ferguson}, {Koekemoer}
  et~al.]{Giavalisco2004}
{Giavalisco} M., {Ferguson} H.~C., {Koekemoer} A.~M., et~al., 2004, \apjl, 600,
  L93

\bibitem[{Gilmour} et~al.(2009){Gilmour}, {Best} \& {Almaini}]{Gilmour2009}
{Gilmour} R., {Best} P., {Almaini} O., 2009, \mnras, 392, 1509

\bibitem[{Grogin} et~al.(2011){Grogin}, {Kocevski}, {Faber} et~al.]{Grogin2011}
{Grogin} N.~A., {Kocevski} D.~D., {Faber} S.~M., et~al., 2011, \apjs, 197, 35

\bibitem[{Haggard} et~al.(2010){Haggard}, {Green}, {Anderson}
  et~al.]{Haggard2010}
{Haggard} D., {Green} P.~J., {Anderson} S.~F., et~al., 2010, \apj, 723, 1447

\bibitem[{H{\"a}ring} \& {Rix}(2004)]{Haering2004}
{H{\"a}ring} N., {Rix} H.-W., 2004, \apjl, 604, L89

\bibitem[{Hasinger} et~al.(2005){Hasinger}, {Miyaji} \&
  {Schmidt}]{Hasinger2005}
{Hasinger} G., {Miyaji} T., {Schmidt} M., 2005, \aap, 441, 417

\bibitem[{Ho} et~al.(2000){Ho}, {Rudnick}, {Rix} et~al.]{ho2000}
{Ho} L.~C., {Rudnick} G., {Rix} H.-W., et~al., 2000, \apj, 541, 120

\bibitem[{Hopkins} et~al.(2008){Hopkins}, {Hernquist}, {Cox} \& {Kere{\v
  s}}]{Hopkins2008}
{Hopkins} P.~F., {Hernquist} L., {Cox} T.~J., {Kere{\v s}} D., 2008, \apjs,
  175, 356

\bibitem[{Jahnke} et~al.(2009){Jahnke}, {Bongiorno}, {Brusa}
  et~al.]{Jahnke2009}
{Jahnke} K., {Bongiorno} A., {Brusa} M., et~al., 2009, \apjl, 706, L215

\bibitem[{Kampczyk} et~al.(2013){Kampczyk}, {Lilly}, {de Ravel}
  et~al.]{Kampczyk2013}
{Kampczyk} P., {Lilly} S.~J., {de Ravel} L., et~al., 2013, \apj, 762, 43

\bibitem[{Kauffmann} et~al.(2003){Kauffmann}, {Heckman}, {Tremonti}
  et~al.]{Kauffmann2003}
{Kauffmann} G., {Heckman} T.~M., {Tremonti} C., et~al., 2003, \mnras, 346, 1055

\bibitem[{Kelly} \& {Shen}(2013)]{Kelly2013}
{Kelly} B.~C., {Shen} Y., 2013, \apj, 764, 45

\bibitem[{Kennicutt}(1998)]{Kennicutt1998}
{Kennicutt} Jr. R.~C., 1998, \apj, 498, 541

\bibitem[{Kenter} et~al.(2005){Kenter}, {Murray}, {Forman} et~al.]{Kenter2005}
{Kenter} A., {Murray} S.~S., {Forman} W.~R., et~al., 2005, \apjs, 161, 9

\bibitem[{Kewley} et~al.(2006){Kewley}, {Groves}, {Kauffmann} \&
  {Heckman}]{Kewley2006}
{Kewley} L.~J., {Groves} B., {Kauffmann} G., {Heckman} T., 2006, \mnras, 372,
  961

\bibitem[{Knobel} et~al.(2013){Knobel}, {Lilly}, {Kova{\v c}}
  et~al.]{Knobel2013}
{Knobel} C., {Lilly} S.~J., {Kova{\v c}} K., et~al., 2013, \apj, 769, 24

\bibitem[{Lehmer} et~al.(2010){Lehmer}, {Alexander}, {Bauer}
  et~al.]{Lehmer2010}
{Lehmer} B.~D., {Alexander} D.~M., {Bauer} F.~E., et~al., 2010, \apj, 724, 559

\bibitem[{Lehmer} et~al.(2007){Lehmer}, {Brandt}, {Alexander}
  et~al.]{Lehmer2007}
{Lehmer} B.~D., {Brandt} W.~N., {Alexander} D.~M., et~al., 2007, \apj, 657, 681

\bibitem[{Lehmer} et~al.(2008){Lehmer}, {Brandt}, {Alexander}
  et~al.]{Lehmer2008}
{Lehmer} B.~D., {Brandt} W.~N., {Alexander} D.~M., et~al., 2008, \apj, 681,
  1163

\bibitem[{Lin} et~al.(2010){Lin}, {Cooper}, {Jian} et~al.]{Lin2010}
{Lin} L., {Cooper} M.~C., {Jian} H.-Y., et~al., 2010, \apj, 718, 1158

\bibitem[{Lusso} et~al.(2012){Lusso}, {Comastri}, {Simmons} et~al.]{Lusso2012}
{Lusso} E., {Comastri} A., {Simmons} B.~D., et~al., 2012, \mnras, 425, 623

\bibitem[{Martel} et~al.(2007){Martel}, {Menanteau}, {Tozzi}, {Ford} \&
  {Infante}]{Martel2007}
{Martel} A.~R., {Menanteau} F., {Tozzi} P., {Ford} H.~C., {Infante} L., 2007,
  \apjs, 168, 19

\bibitem[{Martini} et~al.(2006){Martini}, {Kelson}, {Kim}, {Mulchaey} \&
  {Athey}]{Martini2006}
{Martini} P., {Kelson} D.~D., {Kim} E., {Mulchaey} J.~S., {Athey} A.~A., 2006,
  \apj, 644, 116

\bibitem[{Martini} et~al.(2007){Martini}, {Mulchaey} \& {Kelson}]{Martini2007}
{Martini} P., {Mulchaey} J.~S., {Kelson} D.~D., 2007, \apj, 664, 761

\bibitem[{Merloni} et~al.(2004){Merloni}, {Rudnick} \& {Di
  Matteo}]{Merloni2004}
{Merloni} A., {Rudnick} G., {Di Matteo} T., 2004, \mnras, 354, L37

\bibitem[{O'Sullivan} et~al.(2003){O'Sullivan}, {Ponman} \&
  {Collins}]{OSullivan2003}
{O'Sullivan} E., {Ponman} T.~J., {Collins} R.~S., 2003, \mnras, 340, 1375

\bibitem[{Peng} et~al.(2012){Peng}, {Lilly}, {Renzini} \& {Carollo}]{Peng2012}
{Peng} Y.-j., {Lilly} S.~J., {Renzini} A., {Carollo} M., 2012, \apj, 757, 4

\bibitem[{Pentericci} et~al.(2013){Pentericci}, {Castellano}, {Menci}
  et~al.]{Pentericci2013}
{Pentericci} L., {Castellano} M., {Menci} N., et~al., 2013, \aap, 552, A111

\bibitem[{Pierce} et~al.(2007){Pierce}, {Lotz}, {Laird} et~al.]{Pierce2007}
{Pierce} C.~M., {Lotz} J.~M., {Laird} E.~S., et~al., 2007, \apjl, 660, L19

\bibitem[{Ranalli} et~al.(2003){Ranalli}, {Comastri} \& {Setti}]{Ranalli2003}
{Ranalli} P., {Comastri} A., {Setti} G., 2003, \aap, 399, 39

\bibitem[{Reines} et~al.(2013){Reines}, {Greene} \& {Geha}]{Reines2013}
{Reines} A.~E., {Greene} J.~E., {Geha} M., 2013, \apj, 775, 116

\bibitem[{Richardson} et~al.(2013){Richardson}, {Chatterjee}, {Zheng}, {Myers}
  \& {Hickox}]{Richardson2013}
{Richardson} J., {Chatterjee} S., {Zheng} Z., {Myers} A.~D., {Hickox} R., 2013,
  \apj, 774, 143

\bibitem[{Ruderman} \& {Ebeling}(2005)]{Ruderman2005}
{Ruderman} J.~T., {Ebeling} H., 2005, \apjl, 623, L81

\bibitem[{Sabater} et~al.(2013){Sabater}, {Best} \&
  {Argudo-Fern{\'a}ndez}]{Sabater2013}
{Sabater} J., {Best} P.~N., {Argudo-Fern{\'a}ndez} M., 2013, \mnras, 430, 638

\bibitem[{Sanders} \& {Mirabel}(1996)]{Sanders1996}
{Sanders} D.~B., {Mirabel} I.~F., 1996, \araa, 34, 749

\bibitem[{Sarzi} et~al.(2010){Sarzi}, {Shields}, {Schawinski}
  et~al.]{Sarzi2010}
{Sarzi} M., {Shields} J.~C., {Schawinski} K., et~al., 2010, \mnras, 402, 2187

\bibitem[{Sazonov} \& {Revnivtsev}(2004)]{Sazonov2004}
{Sazonov} S.~Y., {Revnivtsev} M.~G., 2004, \aap, 423, 469

\bibitem[{Schramm} et~al.(2013){Schramm}, {Silverman}, {Greene}
  et~al.]{Schramm2013}
{Schramm} M., {Silverman} J.~D., {Greene} J.~E., et~al., 2013, \apj, 773, 150

\bibitem[{Scoville} et~al.(2007){Scoville}, {Aussel}, {Brusa}
  et~al.]{Scoville2007}
{Scoville} N., {Aussel} H., {Brusa} M., et~al., 2007, \apjs, 172, 1

\bibitem[{S\'ersic}(1968)]{Sersic1968}
{S\'ersic} J.~L., 1968, {Atlas de galaxias australes}

\bibitem[{Shen} et~al.(2007){Shen}, {Mulchaey}, {Raychaudhury}, {Rasmussen} \&
  {Ponman}]{Shen2007}
{Shen} Y., {Mulchaey} J.~S., {Raychaudhury} S., {Rasmussen} J., {Ponman} T.~J.,
  2007, \apjl, 654, L115

\bibitem[{Shen} et~al.(2011){Shen}, {Richards}, {Strauss} et~al.]{Shen2011}
{Shen} Y., {Richards} G.~T., {Strauss} M.~A., et~al., 2011, \apjs, 194, 45

\bibitem[{Shin} et~al.(2012){Shin}, {Ostriker} \& {Ciotti}]{Shin2012}
{Shin} M.-S., {Ostriker} J.~P., {Ciotti} L., 2012, \apj, 745, 13

\bibitem[{Silverman} et~al.(2008){Silverman}, {Green}, {Barkhouse}
  et~al.]{Silverman2008}
{Silverman} J.~D., {Green} P.~J., {Barkhouse} W.~A., et~al., 2008, \apj, 679,
  118

\bibitem[{Silverman} et~al.(2011){Silverman}, {Kampczyk}, {Jahnke}
  et~al.]{Silverman2011}
{Silverman} J.~D., {Kampczyk} P., {Jahnke} K., et~al., 2011, \apj, 743, 2

\bibitem[{Silverman} et~al.(2009{\natexlab{a}}){Silverman}, {Kova{\v c}},
  {Knobel} et~al.]{silverman2009a}
{Silverman} J.~D., {Kova{\v c}} K., {Knobel} C., et~al., 2009{\natexlab{a}},
  \apj, 695, 171

\bibitem[{Silverman} et~al.(2009{\natexlab{b}}){Silverman}, {Lamareille},
  {Maier} et~al.]{Silverman2009b}
{Silverman} J.~D., {Lamareille} F., {Maier} C., et~al., 2009{\natexlab{b}},
  \apj, 696, 396

\bibitem[{Smol{\v c}i{\'c}} et~al.(2011){Smol{\v c}i{\'c}}, {Finoguenov},
  {Zamorani} et~al.]{smolcic2011}
{Smol{\v c}i{\'c}} V., {Finoguenov} A., {Zamorani} G., et~al., 2011, \mnras,
  416, L31

\bibitem[{Strateva} et~al.(2003){Strateva}, {Strauss}, {Hao}
  et~al.]{Strateva2003}
{Strateva} I.~V., {Strauss} M.~A., {Hao} L., et~al., 2003, \aj, 126, 1720

\bibitem[{Tanaka} et~al.(2012){Tanaka}, {Finoguenov}, {Lilly}
  et~al.]{Tanaka2012a}
{Tanaka} M., {Finoguenov} A., {Lilly} S.~J., et~al., 2012, \pasj, 64, 22

\bibitem[{Trump} et~al.(2009){Trump}, {Impey}, {Kelly} et~al.]{Trump2009}
{Trump} J.~R., {Impey} C.~D., {Kelly} B.~C., et~al., 2009, \apj, 700, 49

\bibitem[{Ueda} et~al.(2003){Ueda}, {Akiyama}, {Ohta} \& {Miyaji}]{Ueda2003}
{Ueda} Y., {Akiyama} M., {Ohta} K., {Miyaji} T., 2003, \apj, 598, 886

\bibitem[{Ueda} et~al.(2011){Ueda}, {Hiroi}, {Isobe} et~al.]{Ueda2011}
{Ueda} Y., {Hiroi} K., {Isobe} N., et~al., 2011, \pasj, 63, 937

\bibitem[{van den Bosch} et~al.(2008){van den Bosch}, {Aquino}, {Yang}
  et~al.]{vandenbosch2008}
{van den Bosch} F.~C., {Aquino} D., {Yang} X., et~al., 2008, \mnras, 387, 79

\bibitem[{Weinmann} et~al.(2006){Weinmann}, {van den Bosch}, {Yang}, {Mo},
  {Croton} \& {Moore}]{Weinmann2006}
{Weinmann} S.~M., {van den Bosch} F.~C., {Yang} X., {Mo} H.~J., {Croton} D.~J.,
  {Moore} B., 2006, \mnras, 372, 1161

\bibitem[{Wilman} et~al.(2009){Wilman}, {Oemler}, {Mulchaey}, {McGee}, {Balogh}
  \& {Bower}]{Wilman2009}
{Wilman} D.~J., {Oemler} Jr. A., {Mulchaey} J.~S., {McGee} S.~L., {Balogh}
  M.~L., {Bower} R.~G., 2009, \apj, 692, 298

\bibitem[{Yan} \& {Blanton}(2012)]{Yan2012}
{Yan} R., {Blanton} M.~R., 2012, \apj, 747, 61

\bibitem[{Yang} et~al.(2007){Yang}, {Mo}, {van den Bosch}, {Pasquali}, {Li} \&
  {Barden}]{Yang2007}
{Yang} X., {Mo} H.~J., {van den Bosch} F.~C., {Pasquali} A., {Li} C., {Barden}
  M., 2007, \apj, 671, 153

\end{thebibliography}

 \begin{deluxetable*}{llllllll}
 \tabletypesize{\small}
 \tablecaption{ZENS Galaxy groups with X-ray observations \label{xray_sample}}
\tablehead{\colhead{Name} &\colhead{RA\tablenotemark{a}}&\colhead{DEC\tablenotemark{a}}&\colhead{Redshift\tablenotemark{a}}&\colhead{$N_{H}$\tablenotemark{b}}&\colhead{N$_m\tablenotemark{c}$}&\colhead{$\hat{R}_{200}$\tablenotemark{c}}&\colhead{M$_{group}$\tablenotemark{c}}\\
& \colhead{(J2000)}& \colhead{(J2000)}&&&&\colhead{(Mpc)}&\colhead{($\times 10^{13}M_{\odot})$}}
\startdata
 \hline
\hline
&&&$Chandra$\\
\hline
2PIGG$-$s1571 & 02:37:04.33&-25:23:34.3&0.0568&1.96& 10 &0.501&1.52\\
2PIGG$-$n1610 &09:53:38.23&-05:08:21.4&0.0562&3.85& 10 &0.495&1.45\\
2PIGG$-$n1702 &09:54:30.67&-04:06:03.3&0.0574&3.62& 9 &0.573&2.26\\
2PIGG$-$n1347 &09:59:44.62&-05:16:52.6&0.0521&3.67&10 &0.534&2.90\\
2PIGG$-$n1480 & 10:15:31.91&-05:37:06.9&0.0537&4.50& 13 &0.574&2.26\\
2PIGG$-$n1320 &10:17:55.04&-01:22:53.4&0.0508&4.19& 10 &0.631&3.00\\
2PIGG$-$n1441 &11:18:10.68& -04:27:36.1&0.0531&4.34& 15 &0.658&3.41\\
2PIGG$-$n1381 & 14:28:12.53&-02:31:12.4&0.0522&3.98& 10 &0.468&1.22\\
2PIGG$-$n1598 & 14:35:54.08&-01:16:42.7&0.0560&3.68& 9 &0.606&2.67\\
2PIGG$-$n1746 &14:40:20.07&-03:45:56.2 &0.0585&5.64& 9 &0.516&1.65\\
2PIGG$-$s1752 &22:21:10.68&-26:00:24.6&0.0577&1.65& 11 &0.775&5.60\\
2PIGG$-$s1671 &22:24:00.14& -30:00:17.9&0.0567&1.11& 10 &0.618&2.83\\
\hline
&&&XMM\\
\hline
2PIGG$-$s1520&00:02:01.79&-34:52:55.5&0.0543&1.09&9&0.505&1.55\\
2PIGG$-$s1571&see above\\
2PIGG$-$s1783&22:17:26.33&-36:59:48.1&0.0583&1.18&8&0.741&4.90\\
2PIGG$-$n1606&10:38:49.84&01:48:24.7&0.0561&3.76&7&0.505&1.55\\
2PIGG$-$s1614&22:25:15.88&-25:23:15.4&0.0568&1.72&18&0.746&4.98\\
2PIGG$-$s1471&23:45:01.81&-26:37:26.8&0.0528&1.59&15&0.689&3.91\\
2PIGG$-$n1572&14:25:33.40&-01:30:00.4&0.0550&3.57&19&0.733&4.72
\enddata
\tablenotetext{a}{As reported in \citet{Eke2004a}.}
\tablenotetext{b}{Galactic neutral hydrogen column \citep{Dickey1990}; units of 10$^{20}$ cm$^{-2}$.}
\tablenotetext{a}{Number of galaxy members as reported in \citet{Eke2004a}.}
\tablenotetext{c}{Characteristic group radius and total halo mass as given in \citet{Carollo2013}.}
\end{deluxetable*}

\begin{deluxetable*}{llllll}
\tabletypesize{\small}
\tablecaption{$Chandra$ observation log\label{chandra_log}}
\tablehead{\colhead{Target}&\colhead{RA\tablenotemark{a}}&\colhead{DEC\tablenotemark{a}}&\colhead{Observation}&\colhead{Exposure}&\colhead{OBS}\\
&\colhead{offset (\arcmin)}&\colhead{offset(\arcmin)}&\colhead{date (GMT)}&\colhead{time (ks)}&\colhead{ID}}
\startdata
2PIGG-s1571&+1.86&-2.02&19 Oct 2010&10.06&11613\\
2PIGG-n1610	&-0.45	&-1.73	&22 Jan 2010&	2.62	&	11617\\
&-0.44&-1.74&22 Jan 2010&2.47&11618\\
&-0.43&-1.27&22 Jan 2010&2.47&11619\\
&-0.43&-1.27&22 Jan 2010&2.47&11620\\
2PIGG-n1702&+1.92&-1.51&22 Jan 2010&2.47&11621\\
&+1.91&-1.51&22 Jan 2010&2.47&11622\\
&+1.91&-1.51&22 Jan 2010&2.47&11623\\
&+1.92&-1.51&22 Jan 2010&2.47&11624\\
2PIGG-n1347&+2.19&-4.89&02 Feb 2010&5.17&11625\\
&+3.49&-4.94&08 Feb 2010&5.12&11627\\
2PIGG-n1480&+1.74&+3.46&03 Feb 2010&4.79&11629\\
&+1.93&+4.87&03 Feb 2010&5.18&11631\\
2PIGG-n1320&-0.84&-1.31&28 Jan 2010&2.64&11633\\
&-0.84&-1.31&28 Jan 2010&2.47&11634\\
&-0.84&-1.31&28 Jan 2010&2.47&11635\\
&-0.84&-1.31&28 Jan 2010&2.47&11636\\
2PIGG-n1441&-2.73&+2.28&03 Feb 2010&4.80&11637\\
&-2.89&+1.53&20 Apr 2010&5.17&11639\\
2PIGG-n1381&-3.44&+1.42&07 May 2010&5.06&11641\\
&-4.64&+0.24&24 Dec 2009&2.63&11643\\
&-4.63&+0.29&18 Dec 2009&2.47&11644\\
2PIGG-n1598&-1.54&-5.32&07 May 2010&9.79&11645\\
2PIGG-n1746&-2.35&+4.78&08 May 2010&7.63&11649\\
&-2.39&+7.34&18 Dec 2009&2.55&11652\\
2PIGG-s1752&-0.80&+0.37&09 Sept 2009&5.17&11653\\
&-0.65&-1.80&10 Sept 2009&5.18&11655\\
2PIGG-s1671&-3.23&+0.99&22 July 2010&2.56&11657\\
&-2.98&+0.32&14 Sept 2009&2.67&11658\\
&-3.11&-0.56&19 Sept 2009&2.68&11659\\
&-3.02&-0.54&16 Sept 2009&2.68&11660\\
\enddata
\tablenotetext{a}{Position of the $Chandra$ aim-point given as an offset from the group centers provided in Table~\ref{xray_sample}.}
\end{deluxetable*}

\begin{deluxetable*}{lllllllllll}
\tabletypesize{\small}
\tablecaption{$Chandra$ photometry of ZENS galaxies\label{chandra_xray}}
\tablewidth{0pt}
\tablehead{
\colhead{Name}&\colhead{RA (X-ray)}&\colhead{DEC (X-ray)}&\colhead{Off-axis}&&\colhead{Counts}&&\colhead{$f_X$\tablenotemark{a}}&\colhead{$L_X$\tablenotemark{b}}&\colhead{Class}\\
&\colhead{(J2000)}&\colhead{(J2000)}&\colhead{angle ($\arcmin$)}&\colhead{B}&\colhead{S}&\colhead{H}&&&}
\startdata
s1571\_10&02:36:55.95&-25:28:39.4&4.99&$6.1_{-2.4}^{+3.1}$&$3.8_{-1.1}^{+2.4}$&$2.3_{-1.5}^{+2.2}$&-14.2&40.7&AGN\\
s1671\_4&22:23:54.06&-30:00:46.3&2.41&$10.2_{-3.9}^{+3.9}$&$8.8_{-3.2}^{+3.2}$&$1.4_{-1.4}^{+2.3}$&-13.9&40.9&Galaxy\\
s1752\_4&22:21:10.59&-26:00:24.6&1.40&$11.5_{-3.9}^{+3.9}$&$9.8_{-3.1}^{+3.6}$&$1.8_{-1.6}^{+2.1}$&-13.9&41.0&AGN\\
n1347\_3\tablenotemark{c}&09:59:44.04&-05:22:04.9&3.46&$12.0_{-4.0}^{+4.6}$&$5.3_{-2.4}^{+3.0}$&$6.8_{-3.0}^{+3.7}$&-13.8&41.0&Galaxy\\
n1381\_3&14:28:06.04&-02:31:27.32&2.86&$4.8_{-2.8}^{+3.3}$&$2.8_{-1.8}^{+2.3}$&$2.0_{-1.8}^{+2.2}$&-14.4&40.6&AGN\\
n1598\_5&14:36:16.89&-01:23:13.65&8.15&$9.1_{-3.2}^{+3.8}$&$5.8_{-2.4}^{+3.0}$&$3.2_{-2.0}^{+2.6}$&-13.9&40.9&AGN\\
n1702\_3\tablenotemark{d}&&&4.72&$4.6_{-3.3}^{+2.7}$&$0.7_{-0.7}^{+1.8}$&$3.9_{-2.4}^{+2.9}$&-14.2&40.5&AGN\\
\hline
n1381\_add&14:28:08.90&-02:31:24.68&3.18&$80.1_{-9.0}^{+9.1}$&$15.3_{-3.7}^{+4.3}$&$64.8_{-8.1}^{+8.2}$&-13.0&41.8&AGN\\
\enddata
\tablenotetext{a}{The log of the flux in units erg cm$^{-2}$ s$^{-1}$ in the 0.5-8.0 keV band.}
\tablenotetext{b}{The log of the luminosity in units erg s$^{-1}$ in the 0.5-8.0 keV band.}
\tablenotetext{c}{X-ray emission is slightly extended and X-ray source is not found by wavdetect; given position indicates peak X-ray emission}
\tablenotetext{d}{No position is given since source is not detected by wavdetect with a significance above our threshold.  The optical position was used for count extraction.}
\end{deluxetable*}

\begin{deluxetable*}{lllllllll}
\tabletypesize{\small}
\tablecaption{XMM-$Newton$ photometry of ZENS galaxies\label{xmm_xray}}
\tablewidth{0pt}
\tablehead{
\colhead{Name}&\colhead{RA (X-ray)}&\colhead{DEC (X-ray)}&\colhead{Count rate\tablenotemark{a}}&\colhead{Count rate\tablenotemark{a}}&\colhead{Flux\tablenotemark{b}}&\colhead{$L_X$\tablenotemark{c}}&\colhead{Class}\\
&\colhead{(J2000)}&\colhead{(J2000)}&\colhead{S}&\colhead{H}&\colhead{B}&\colhead{B}}
\startdata
s1520\_4&00:02:01.6&-34:52:56&6.8$\pm$1.7&$<1.6$&-13.9&41.0&AGN\\
s1571\_8&02:37:04.4&-25:23:35&7.7$\pm$1.3&$<1.1$&-13.8&41.1&Galaxy\\
s1783\_2&22:17:26.3&-36:59:50&8.2$\pm$1.1&2.2$\pm$0.6&-13.8&41.0&AGN\\
s1783\_3&22:17:20.9&-36:58:22&5.8$\pm$0.8&$<1.2$&-14.0&41.0&AGN\\
s1783\_4&22:17:19.3&-36:56:50&7.0$\pm$0.9&1.9$\pm$0.6&-13.9&41.0&AGN\\
s1783\_7&22:17:06.2&-36:56:54&3.0$\pm$0.8&2.2$\pm$0.6&-14.2&40.8&AGN\\
n1606\_3&10:38:45.4&+01:46:45&3.9$\pm$1.1&$<1.5$&-14.1&40.8&AGN\\
n1606\_5&10:38:41.2&+01:43:29&3.2$\pm$1.0&$<1.6$&-14.2&40.7&Galaxy\\
s1614\_10&22:25:19.0&-25:24:23&8.7$\pm$2.7&$<3.6$&-13.8&41.1&AGN\\
s1471\_4&23:45:44.0&-26:43:07&107.2$\pm$9.8&33.1$\pm$5.2&-12.7&42.1&AGN\\
s1471\_5&23:45:34.4&-26:43:35&12.0$\pm$3.4&43.9$\pm5.4$&-13.1&41.7&AGN\\
s1471\_11&23:45:05.5&-26:40:47&10.8$\pm$2.0&$<1.8$&-13.7&41.1&Galaxy\\
s1471\_12&23:45:01.9&-26:37:23&4.7$\pm$1.3&$<2.2$&-14.0&40.8&Galaxy\\
n1572\_13&14:25:55.0&-01:28:15&6.4$\pm$1.7&$<2.6$&-13.9&40.9&AGN\\
\hline
\enddata
\tablenotetext{a}{units $10^{-4}$ s$^{-1}$ in either the observed soft (S: 0.5-2.0 keV) or hard (H: 2.0-7.5 keV) band}
\tablenotetext{b}{The log of the flux in units erg cm$^{-2}$ s$^{-1}$ in the 0.5-8.0 keV band (see text for details)}
\tablenotetext{c}{The log luminosity in units erg s$^{-1}$ in the 0.5-8.0 keV band (see text for details)}
\end{deluxetable*}

\clearpage

\begin{deluxetable*}{llllllllllll}
\tabletypesize{\small}
\tablecaption{Optical properties of X-ray sources associated with ZENS galaxies I. \label{opt_table1}}
\tablewidth{0pt}
\tablehead{\colhead{Name}&\colhead{RA (opt)}&\colhead{DEC (opt)}&\colhead{Redshift}&\colhead{B\tablenotemark{a}}&\colhead{I\tablenotemark{a}}&\colhead{$K_s$\tablenotemark{b}}&\colhead{$M_{galaxy}$\tablenotemark{c}}&\colhead{$SFR$}&\colhead{Morph.}\\
&\colhead{(J2000)}&\colhead{(J2000)}&&&&&\colhead{($\times10^{10} M_{\odot})$}&\colhead{(M$_{\odot}$ yr$^{-1}$)}&\colhead{type\tablenotemark{d}}}
\startdata
s1571\_10&02:36:55.96&-25:28:39.9&0.0575&19.2&18.2&18.2&0.077&1.7&5\\
s1671\_4&22:23:54.12&-30:00:47.7&0.0568&15.5&14.1&13.4&14.8&0.03&0\\
s1752\_4&22:21:10.69&-26:00:24.8&0.0569&15.3&13.8&13.1&31.3&0.09&1\\
n1347\_3&09:59:44.18&-05:22:01.2&0.0526&14.8&13.3&12.5&28.2&0.08&0\\
n1381\_3&14:28:06.07&-02:31:24.5&0.0522&17.3&16.6&16.1&0.75&0.86&4\\
n1598\_5&14:36:16.84&-01:23:13.3&0.0556&15.6&14.5&13.8&5.92&15.2&3\\
n1702\_3&09:54:57.08&-04:05:14.6&0.0574&19.0&18.1&---&0.15&0.22&4\\
\hline
s1520\_4&00:30:27.00&-34:52:55.7&0.0548&16.2&14.8&14.1&5.56&0.04&1\\
s1571\_8&02:37:04.34&-25:23:34.5&0.0570&15.8&14.4&13.6&16.7&0.13&0\\
s1783\_2&22:17:26.34&-36:59:48.3&0.0583&16.5&15.0&14.0&12.5&0.12&0\\
s1783\_3&22:17:20.86&-36:58:23.0&0.0588&17.1&15.6&14.8&3.73&0.31&3\\
s1783\_4&22:17:19.53&-36:56:45.3&0.0586&16.6&15.2&14.7&5.93&5.93e-4&3\\
s1783\_7&22:17:06.24&-36:56:51.3&0.0587&15.0&13.7&13.1&21.7&8.56&3\\
n1606\_3&10:38:45.39&+01:46:43.4& 0.0566&16.3&15.2&14.6&3.05&6.45&3\\
n1606\_5&10:38:41.52&+01:43:28.1&0.0555&16.5&15.8&15.0&0.75&2.94&3\\
s1614\_10&22:25:19.02&-25:24:26.70&0.0586&16.1&15.0&14.3&2.39&17.56&2\\
s1471\_4&23:45:43.81&-26:43:10.0&0.0515&15.8&15.0&14.6&2.27&4.12&4\\
s1471\_5&23:45:34.24&-26:43:36.6&0.0514&17.1&15.8&15.4&2.40&4.7e-3&2\\
s1471\_11&23:45:05.72&-26:40:48.1&0.0522&15.8&14.3&---&20.2&2.02e-3&3\\
s1471\_12&23:45:01.82&-26:37:27.0&0.0520&15.9&14.4&13.7&23.40&1.67&2\\
n1572\_13&14:25:55.11&-01:28:18.1&0.0550&16.6&15.2&14.8&9.53&$<10^{-4}$&0\\
\hline
&&&&$g_{sdss}$&$i_{sdss}$&$K_s$\tablenotemark{a}\\
\hline
n1381\_add\tablenotemark{e}&14:28:08.89&-02:31:24.8&0.0520&18.6&17.4&15.1&0.69&0.12&0\\
\enddata
\tablenotetext{a}{Petrosian apparent magnitude; rest-frame}
\tablenotetext{b}{2MASS apparent magnitude; rest-frame}
\tablenotetext{c}{Stellar mass}
\tablenotetext{d}{ZENS morphological type: 0=elliptical, 1=S0, 2=Bulge-dominated spiral; 3=Intermediate spiral; 4=Disk-dominated spiral; 5=irregular}
\tablenotetext{e}{SDSS J142808.89-023124.8}
\end{deluxetable*}

\begin{deluxetable*}{lcllllllll}
\tabletypesize{\small}
\tablecaption{Optical properties of X-ray sources associated with ZENS galaxies II.\label{opt_table2}}
\tablewidth{0pt}
\tablehead{\colhead{Name}&\colhead{Central\tablenotemark{a}}&\colhead{$R/{\hat R}_{200}\tablenotemark{b}$}&\colhead{$n_{sersic}$\tablenotemark{c}}&\colhead{$n_{sersic}$\tablenotemark{c}}&\colhead{$r_{1/2}^{total}$\tablenotemark{~c,d}}&\colhead{$r_{1/2}^{bulge}$\tablenotemark{~c,e}}&\colhead{B/T\tablenotemark{c,f,g}}&\colhead{h: disk scale}\\
&\colhead{flag}&&\colhead{(total)}&\colhead{(bulge)}&\colhead{(kpc)}&\colhead{(kpc)}&&\colhead{length (kpc)\tablenotemark{c,g}}}
\startdata
s1571\_10 & 0 & 0.749 & \nodata & \nodata & 4.21 &\nodata& \nodata&\nodata\\
s1671\_4 & 1 & 0 & 3.92 & \nodata & 8.10 &\nodata& \nodata & \nodata \\
s1752\_4 & 1 & 0 & 3.34 & $2.82_{-0.07}^{+0.06} $& 7.39 &$3.92_{-0.11}^{+0.12}$&$0.623_{-0.014}^{+0.014}$&$5.869_{-0.142}^{+0.153}$\\
n1347\_3 & 1 & 0 & 5.64 & \nodata& 16.37 &\nodata&\nodata &  \nodata\\
n1381\_3 & 0 & 0.241 & 1.10 & $2.69_{-0.07}^{+0.13}$ & 2.90 &$0.03_{-0.03}^{+0.07}$&$0.018_{-0.002}^{+0.004}$ & $1.790_{-0.019}^{+0.017}$\\
n1598\_5 & 1 & 0 & \nodata & $3.63_{-0.06}^{+0.04}$ & 10.50 &$5.16_{-0.23}^{+0.05}$&$0.314_{-0.009}^{+0.003}$ & $7.384_{-0.136}^{+0.024}$\\
n1702\_3 & 0 & 0.801 & 1.49 & $5.37_{-0.63}^{+0.56}$ & 2.23 &$1.44_{-0.18}^{+0.19}$&$0.183_{-0.027}^{+0.032}$&$1.441_{-0.055}^{+0.054}$ \\
\hline
s1520\_4 & 1 & 0 & 6.33 & $3.23_{-0.06}^{+0.07}$ & 17.63 &$2.47_{-0.05}^{+0.05}$&$0.497_{-0.006}^{+0.007}$ &$7.542_{-0.244}^{+0.089}$\\
s1571\_8 & 1 & 0 & 6.82 & \nodata& 14.12 &\nodata&\nodata & \nodata \\
s1783\_2 & 0 & 0.465 & 3.01 & \nodata & 4.51 &\nodata&\nodata&  \nodata\\
s1783\_3 & 0 & 0.308 & 9.28 & $2.29_{-0.13}^{+0.15}$ & 7.63 &$0.64_{-0.02}^{+0.02}$&$0.327_{-0.008}^{+0.008}$ & $2.872_{-0.031}^{+0.028}$\\
s1783\_4 & 0 & 0.248 & 2.64 & $4.20_{-0.19}^{+0.23}$ & 4.95 & $3.46_{-0.23}^{+0.13}$&$0.399_{-0.016}^{+0.011} $&$3.305_{-0.040}^{+0.067}$\\
s1783\_7 & 1 & 0 & 3.86 & $4.96_{-0.48}^{+0.15}$ & 14.0 &$3.53_{-0.39}^{+0.59}$&$0.268_{-0.021}^{+0.044} $&$6.614_{-0.094}^{+0.107}$\\
n1606\_3 & 0 & 0.278 & 2.09 & $9.99_{-1.17}^{+0.01}$ & 3.54 & $1.33_{-0.54}^{+0.98}$&$0.202_{-0.033}^{+0.057}$ & $2.115_{-0.015}^{+0.027}$ \\
n1606\_5 & 0 & 0.736 & 2.15 & $1.70_{-0.29}^{+0.13}$ & 4.23 &$2.62_{-0.12}^{+0.34}$&$0.215_{-0.013}^{+0.024}$ & $3.526_{-0.179}^{+0.066}$\\
s1614\_10 & 0 & 0.253 & 3.02 & \nodata& 5.95 &\nodata&\nodata& \nodata \\
s1471\_4 & 0 & 1.104 & 1.96 & $4.48_{-0.24}^{+0.19}$ & 6.59 & $3.18_{-0.26}^{+0.06}$&$0.137_{-0.020}^{+0.007}$ &  $4.071_{-0.173}^{+0.019}$\\
s1471\_5 & 0 & 0.955 & 3.15 & $3.50_{-0.19}^{+0.33}$ & 1.64 & $0.95_{-0.11}^{+0.28}$ &$0.518_{-0.039}^{+0.085}$ &  $1.274_{-0.081}^{+0.029}$\\
s1471\_11 & 0 & 0.348 & \nodata& $1.39_{-0.09}^{+0.07}$ & 5.58 &$1.18_{-0.04}^{+0.05}$&$0.434_{-0.023}^{+0.009}$ &$5.412_{-0.162}^{+0.169}$\\
s1471\_12 & 1 & 0 & 3.46 & $2.84_{-0.04}^{+0.04}$ & 7.48 &$3.60_{-0.10}^{+0.10}$&$0.776_{-0.008}^{+0.008}$ & $7.946_{-0.310}^{+0.310}$\\
n1572\_13 & 1 & 0 & 4.47 & \nodata& 3.49 &\nodata&\nodata& \nodata \\
\enddata
\tablenotetext{a}{Central galaxy =1 and satellite=0}
\tablenotetext{b}{Distance of the galaxy to the group center in units of ${\hat R}_{200}$ as defined in \citet{Carollo2013}.  A value of zero denotes a central galaxy}
\tablenotetext{c}{based on the I-band imaging}
\tablenotetext{d}{Half-light radius of the total galaxy light profile.}
\tablenotetext{e}{Half-light radius of the bulge component of galaxy light profile.}
\tablenotetext{f}{Ratio of the bulge-to-total luminosity.}
\tablenotetext{g}{Formal errors based on GIM2D; larger systematic errors are discussed in \citet{Cibinel2013}}
\end{deluxetable*}

\begin{deluxetable*}{lllll}
\tabletypesize{\small}
\tablecaption{AGN statistics in ZENS galaxies\label{table:rank}}
\tablewidth{0pt}
\tablehead{
\colhead{Type}&Mass range&\colhead{\# of galaxies}&\colhead{\#with AGNs}&\colhead{AGN fraction}
}
\startdata
Central&10.4-11.6&14&5&$0.36_{-0.11}^{+0.14}$\\
&10.4-11.0&7&3&$0.42_{-0.14}^{+0.19}$\\
&11.0-11.6&7&2&$0.28_{-0.11}^{+0.20}$\\
Satellite&10.4-11.6&30&4&$0.13_{-0.04}^{+0.09}$\\
&10.4-11.0&28&3&$0.11_{-0.04}^{+0.08}$\\
&11.0-11.6&2&1&$0.50_{-0.25}^{+0.25}$\\
\enddata
\end{deluxetable*}

\end{document}